\documentclass[11pt,notoc]{JHEP3}
\usepackage{graphicx}
\usepackage{amssymb}
\usepackage{amsmath}
\usepackage{epstopdf}
\usepackage{latexsym}
\usepackage[all]{xy}

\newcommand{\be}{\begin{equation}}
\newcommand{\beq}[1]{\begin{equation}\label{#1}}
\newcommand{\eq}[1]{Eq.~(\ref{#1})}

\newcommand{\ee}{\end{equation}}
\newcommand{\eeq}{\end{equation}}
\newcommand{\bea}{\begin{eqnarray}}
\newcommand{\eea}{\end{eqnarray}}
\newcommand{\comment}[1]{}
\newcommand{\TeV}{~\mathrm{TeV}}

\newcommand{\Pl}{\mathrm{Pl}}
\newcommand{\lsim}{\!\mathrel{\hbox{\rlap{\lower.55ex \hbox{$\sim$}} \kern-.34em \raise.4ex \hbox{$<$}}}}
\newcommand{\gsim}{\!\mathrel{\hbox{\rlap{\lower.55ex \hbox{$\sim$}} \kern-.34em \raise.4ex \hbox{$>$}}}}

\DeclareMathOperator{\tr}{tr}

\newcommand{\doublerightxyarrow}{\ar@{-}[rr] |-{\SelectTips{eu}{}\object@{>}}}

\newcommand{\noarrow}{\ar@{-}[r] |-{\SelectTips{eu}{}\object@{}}}

\newcommand{\rightxyarrow}{\ar@{-}[r] |-{\SelectTips{eu}{}\object@{>}}}
\newcommand{\rightdownxyarrow}{\ar@{-}[rd] |-{\SelectTips{eu}{}\object@{>}}}
\newcommand{\downxyarrow}{\ar@{-}[d] |-{\SelectTips{eu}{}\object@{>}}}
\newcommand{\rightupxyarrow}{\ar@{-}[ru] |-{\SelectTips{eu}{}\object@{>}}}
\newcommand{\upxyarrow}{\ar@{-}[u] |-{\SelectTips{eu}{}\object@{>}}}
\newcommand{\leftupxyarrow}{\ar@{-}[lu] |-{\SelectTips{eu}{}\object@{>}}}
\newcommand{\leftdownxyarrow}{\ar@{-}[ld] |-{\SelectTips{eu}{}\object@{>}}}
\newcommand{\leftxyarrow}{\ar@{-}[l] |-{\SelectTips{eu}{}\object@{>}}}

\newcommand{\longdashedrightxyarrow}{\ar@{--}[rr] |-{\SelectTips{eu}{}\object@{>}}}
\newcommand{\longdashedrightdownxyarrow}{\ar@{--}[rdrd] |-{\SelectTips{eu}{}\object@{>}}}
\newcommand{\longdasheddownxyarrow}{\ar@{--}[dd] |-{\SelectTips{eu}{}\object@{>}}}
\newcommand{\longdashedrightupxyarrow}{\ar@{--}[ruru] |-{\SelectTips{eu}{}\object@{>}}}
\newcommand{\longdashedupxyarrow}{\ar@{--}[uu] |-{\SelectTips{eu}{}\object@{>}}}
\newcommand{\longdashedleftupxyarrow}{\ar@{--}[lulu] |-{\SelectTips{eu}{}\object@{>}}}
\newcommand{\longdashedleftdownxyarrow}{\ar@{--}[ldld] |-{\SelectTips{eu}{}\object@{>}}}
\newcommand{\longdashedleftxyarrow}{\ar@{--}[ll] |-{\SelectTips{eu}{}\object@{>}}}

\newcommand{\altleftxyarrow}{\ar@{-}[r] |-{\SelectTips{eu}{}\object@{<}}}

\newcommand\openone{\leavevmode\hbox{\small1\normalsize\kern-.33em1}}

\title{Little M-theory}
\author{Hsin-Chia Cheng \\ Department of Physics, University of California, Davis, CA 95616 \\ E-mail: \email{cheng@physics.ucdavis.edu}}
\author{Jesse Thaler \\ Jefferson Physical Laboratory, Harvard University, Cambridge, MA 02143 \\ Department of Physics, University of California, Berkeley, CA 94720 \\
Theoretical Physics Group, Lawrence Berkeley National Laboratory, Berkeley, CA 94720 \\
E-mail: \email{jthaler@jthaler.net}}
\author{Lian-Tao Wang \\ Jefferson Physical Laboratory, Harvard University, Cambridge, MA 02143 \\ E-mail: \email{liantaow@schwinger.harvard.edu}}

\abstract{
Using the language of theory space, \emph{i.e.}\ moose models, we develop a
unified framework for studying composite Higgs models at the LHC.  This framework---denoted little M-theory---is
conveniently described by a theoretically consistent three-site moose
diagram which implements minimal flavor and isospin violation.  By taking different limits of the couplings, one can interpolate between simple group-like and minimal moose-like models
with and without $T$-parity.  In this way, little M-theory reveals a large model
space for composite Higgs theories. We argue that this framework is suitable as a starting point for a comprehensive study of composite Higgs scenarios. The rich collider
phenomenology of this framework is briefly discussed. 
}

\begin{document}

\section{Motivation}

If there is a natural solution to the hierarchy problem, then the data
from the Large Hadron Collider (LHC) will be spectacular.  Almost all
natural theories predict new colored states around the TeV scale to
cancel the quadratically divergent top contribution to the Higgs
potential, and at a $pp$ collider, there is a large cross section for
producing these new states.   But while we will certainly know
\emph{whether} there is new physics at the TeV scale, it may be more
difficult to know \emph{what} new physics we are actually producing.
As the commissioning of the LHC draws near, it is important to
simulate a large number of models to determine which experimental
observables can best distinguish different possibilities for physics
beyond the standard model.


Many scenarios for stabilizing the electroweak scale have been
proposed.  Besides low energy
supersymmetry (SUSY), there are various classes of non-SUSY theories
in the literature:  technicolor
\cite{Weinberg:1975gm,Susskind:1978ms}, top color \cite{Hill:1991at},
Higgsless
\cite{Csaki:2003dt}, Universal Extra Dimensions (UED)
\cite{Arkani-Hamed:2000hv,Appelquist:2000nn}, little Higgs
\cite{Arkani-Hamed:2001nc,Arkani-Hamed:2002pa}, holographic Higgs
\cite{Contino:2003ve,Agashe:2004rs}, twin Higgs \cite{Chacko:2005pe},
and so on.  Within each
class, there are a large number of variants that are in principle
distinguishable given enough experimental data, and each variant has a
set of adjustable parameters.   It is challenging task to decide
whether any specific model deserves a detailed study of its collider
signals. Moreover, it is not clear whether any particular model should
be treated as unique in its class or part of a bigger continuous model
space.

In this paper, we attempt to develop a unified framework for
describing a large class of non-SUSY models, with the goal of
mimicking the current situation for supersymmetric theories.  Though
there are a lot of SUSY-breaking mechanisms which predict different
low energy spectra, we have general low energy
Lagrangians---\emph{i.e.}\ the MSSM \cite{Dimopoulos:1981zb} and its
extensions---whose
parameter space interpolates among various SUSY-breaking
models. Therefore, at the LHC we need not conduct specific searches
for any particular SUSY-breaking mechanism. Rather, we can
attempt to measure the parameters of a general Lagrangian, and such
measurements will hopefully lead us towards a particular model.

The key observation that makes non-SUSY unified frameworks possible is
that the low energy physics of many seemingly dissimilar theories can
actually be different limits of the
same theory once we integrate out degrees of freedom inaccessible at
the LHC.  Because these frameworks are most simply described by moose
diagrams \cite{Georgi:1985hf}, we call them ``M-theories''.  As we will see,
most non-SUSY theories have an associated moose, and different non-SUSY
theories are often associated with the \emph{same} moose, so by studying the
LHC phenomenology of a single M-theory, one can simultaneously explore
many different non-SUSY solutions to the hierarchy problem.   Regardless of
whether an extra
dimension is flat \cite{ADD} or warped \cite{RS1,RS2}, or whether a little
Higgs comes from a
simple group \cite{Kaplan:2003uc,Schmaltz:2004de} or a minimal moose
\cite{Arkani-Hamed:2002qx,Chang:2003un}, there is a single low energy
effective description relevant for the LHC.  The utility of such a general
framework is not just that it interpolates among various known models; it
also reveals the existence of a larger model space with richer structure
than any of the known limits.

There are two facts---one theoretical, one experimental---that make M-theories relevant for the LHC.  On the theory side, it is important that
moose diagrams are general enough to approximate the low energy physics of
known non-SUSY theories.  Almost all non-SUSY theories for physics beyond
the standard model are either (a) based on extra dimensions, (b) based on
moose diagrams, (c)
well-approximated by extra dimensions using the AdS/CFT correspondence
\cite{Maldacena:1997re,Gubser:1998bc,Witten:1998qj,Arkani-Hamed:2000ds,Rattazzi:2000hs}, or (d) well-approximated by mooses using the technique of little
technicolor \cite{Thaler:2005kr}.  With the help of deconstruction
\cite{Arkani-Hamed:2001ca,Hill:2000mu}, we can indeed combine all four
possibilities into M-theories, moose diagrams whose various limits reproduce
the important features of different underlying non-SUSY theories.

On the experimental side, it is important that 
a ``low energy'' moose description can be justified at the energy
scales accessible to the LHC.
Otherwise, a moose would fail to capture interesting ``high energy'' LHC
signatures and a more complete theory would be needed.   In many non-SUSY
models, there is a layer of weakly coupled physics between the electroweak
scale and a cut-off $\Lambda$.  The hierarchy between $M_\Pl$ and $\Lambda$
is either taken care of by some
strong dynamics or left to an unspecified UV completion.  As long as $
\Lambda$ is heavy enough, the LHC cannot probe the underlying UV
model directly, and a weakly coupled low energy effective description will
suffice.  Though the center of mass energy of the LHC is $14 \TeV$, the
discovery reach is typically smaller by about factor of 3 or 4, so
$\Lambda$ need only be a few TeV to safely use an M-theory
approximation.  This does mean, 
however, that M-theories will not be as useful for describing theories like
technicolor, where the scale of strong dynamics is expected to be within the
reach of the LHC.

A straightforward method to construct a non-SUSY unified framework
would be to use a purely bottom-up approach motivated by our UV ignorance.
Apart from high energy modes, various non-SUSY models yield largely similar
low energy spectra.  Therefore, one could simply write down a theory that
includes only the layer of new physics states
between the electroweak scale and $\Lambda$, which usually fill out simple
representations of $SU(2)_L \times U(1)_Y$.  There could be extra
broken $U(1)$s or other extensions of the electroweak gauge
symmetries, but with certain assumptions of minimality, it is not hard to
write down a generic Lagrangian which parametrizes the interactions of all
of those states.

In this study, we seek to go beyond such a bottom-up approach and
incorporate lessons from the past several decades of non-SUSY model
building.  In the standard model, we minimally expect new
physics at $\Lambda \sim 4 \pi v_{\rm EW}$ to unitarize $W$-$W$
scattering.  However, if we naively use $4 \pi v_{\rm EW}$ to set the
size of higher dimension operators in the standard model, then we find
too large corrections to electroweak observables constrained by precision measurements. 
This tension at the percent-level is known as the little hierarchy
problem \cite{Barbieri:2000gf}, which has
received a lot of attention in the post-LEP era. It is therefore
particularly well-motivated to study models that attempt to explain the
separation between the electroweak scale and a cut-off $\Lambda$ at around
$10 \TeV$, the
scale of new physics suggested by precision electroweak measurements
\cite{Barbieri:2004qk}.  Like the bottom-up approach, we
start by writing down an effective theory beneath the cut-off scale, but we
will try to preserve as many of features of realistic theories as we can in
order to focus our attention on preferred regions of model space.  In
particular, our framework allows for $T$-parity
\cite{Cheng:2003ju,Cheng:2004yc}, a $\mathbf{Z}_2$ symmetry that
generically protects precision observables.

From the point of view of the little hierarchy, one of the most
interesting non-SUSY scenarios to be probed at the LHC is a composite Higgs
\cite{Kaplan:1983fs,Kaplan:1983sm}.  Generally speaking, these are
models where there is a Goldstone mode with the quantum numbers of a
Higgs doublet and where same-statistics partners cancel divergent
contributions to the Higgs potential.   In composite Higgs theories (as
opposed to technicolor or Higgsless theories) there is a range of energies
in which $W$-$W$ scattering is
unitarized by a weakly coupled Higgs doublet, and only at a higher
scale $\Lambda \sim 4 \pi f_{\rm eff}$ does one see that the doublet
is accompanied by additional, possibly strongly coupled, high energy
modes.  As long as there is a mechanism to guarantee $ f_{\rm eff} \gg v_{\rm EW}$, then not only will the na\"{\i}ve corrections to precision
electroweak observables be suppressed, but the scale $\Lambda$ will also be
beyond the reach of the LHC, justifying an M-theory description.

Composite Higgs theories can be classified according to the quantum numbers
of the new
heavy modes that regulate higgs-gauge loops.  In minimal moose-like
theories, the electroweak gauge group is doubled, yielding massive
$W'$ partners.  In simple group-like theories, the electroweak gauge
group is embedded in a larger gauge group, yielding massive $X/Y$
(off-diagonal) gauge boson
partners.  For example, most collider studies to date have focused on
the littlest Higgs \cite{Arkani-Hamed:2002qy}, which is a minimal moose-like
theory because the
fundamental gauge group contains two copies of $SU(2)$.   Holographic
composite Higgs models are a hybrid scenario with both $W'$ and $X/Y$
states.  Here, we construct a ``little M-theory'' suitable for
collider studies that interpolates between both choices for the new
heavy spin-1 modes, providing a single phenomenological model where many
different LHC signatures can be explored.

While the term ``M-theory'' \cite{Schwarz:1995jq} suggests the existence of
a unique description of non-SUSY LHC physics, there are in fact many
different M-theories just as there are many different SUSY extensions of the
standard model.  In general, there are two orthogonal directions one could
explore in the model space of mooses.  In this paper, we focus on the
different mechanisms one can employ for canceling quadratic divergences,
studying various limits of one underlying moose.   However, the requirement
of a composite Higgs does not fix the symmetry structure and its breaking
pattern, so one could also explore models with different global and gauge
symmetries.   In the spirit of the MSSM, we choose the minimal symmetry
structure which still allows for custodial $SU(2)$.  We argue that this is a
useful framework which contains generic phenomenology.  It is
straightforward to adjust the symmetry structure of the moose if we
experimentally discover more or less exotics.

To summarize, the new physics which will be probed by the
LHC can be described two broad categories of models: SUSY theories and
theory space theories.   Just as the MSSM is a interesting example of
a SUSY theory, little M-theory is an interesting moose model that can
interpolate between many different ultraviolet models.  Though it is
indeed possible to deform this little M-theory into a UED or Higgsless
model, we will stay in the composite Higgs limit in order to keep the
little M-theory Lagrangian as simple as possible.   We comment on the
implications of such deformations in the conclusion.

In the next section, we review the low energy equivalence between
different types of non-SUSY models and present a toy little M-theory
that interpolates between three known composite Higgs models based on
very different starting assumptions.  Readers interested in the actual
model can skip directly to Section \ref{sec:sp4so4} where we present a
theoretically
consistent little M-theory based on the coset space $Sp(4)/SO(4)$.
This moose is suitable for collider studies and has many adjustable
parameters to deform the spectrum and decay modes.  
In Section 4, we discuss experimental constraints and preferred region of the parameter space, and describe several familiar limits of little
M-theory. 
Interesting new features of the phenomenology 
as well as open questions are commented in Section 5.
Conclusions and the possibility of extending the little M-theory approach to other classes of non-SUSY theories are contained in Section 6.

\section{Known Mooses and Little M-theories}
\label{sec:knownmooses}


As already mentioned, there are two observations that justify the use
of little M-theories for describing LHC phenomenology.  The first is
that different ultraviolet theories can have the same low energy
physics.  A classic example  
of this is a KK tower.  If we deconstruct
a (non-gravitational) extra dimension \cite{Arkani-Hamed:2001ca,Hill:2000mu}, then the first $n$ KK
modes are well approximated by an $n$-site moose diagram.  We
can improve the approximation either by adding sites or by
introducing non-local interactions in theory space.   In this way, an extra dimensional theory and an $n$-site moose theory have nearly identical LHC phenomenology as long as only the first $n$
KK modes are kinematically accessible at the LHC.  Moreover, both flat
and warped extra dimensions can be described by the same $n$-site
moose, the only difference being the values of the gauge couplings and
decay constants on the sites and links
\cite{Cheng:2001nh,Randall:2002qr}.   In the next subsection, we
review low energy equivalences in the context of electroweak physics
and show why mooses are a convenient way to encode infrared degrees of
freedom.    

The second justification for using little M-theories is that if we
take the little hierarchy problem seriously, then we expect a
hierarchical separation between $v_{\rm EW}$ and $f_{\rm eff}$.   In
this way, the scale of possible strong dynamics $\Lambda \sim 4 \pi
f_{\rm eff}$ is beyond the reach of the LHC, and a weakly coupled
moose description will suffice.\footnote{In the context of both
  composite Higgs and technicolor theories, there have been attempts
  to bring the ratio $f_{\rm eff} / v_{\rm EW}$ closer to 1 while
  evading the na\"{\i}ve bounds from precision electroweak
  measurements (see \emph{e.g.}\
  \cite{Luty:2004ye,Agashe:2005dk,Agashe:2006at}).  If
  strong dynamics is seen at the LHC, then a moose description will
  act like a ``techni-QCD'' chiral Lagrangian.}   From a model
building perspective, one would like some symmetry reason to guarantee
this little hierarchy.  Indeed, the novel structure of the Higgs
potential is the \emph{raison d'\^{e}tre} for little Higgs theories,
in that these theories exhibit a parametric separation between $v_{\rm
  EW}$ and $f_{\rm eff}$ compared to generic composite models.  From
the point of view of the LHC, though, the origin of the Higgs
potential has little impact on collider signatures.  For example, the
minimal moose contains extra link fields and plaquette operators to
generate a large enough Higgs quartic without introducing a large
Higgs mass, but this extended Higgs sector just introduces new heavy
states with no generic pattern.   Thus, in both the toy example in
Section \ref{sec:toy} and the complete theory in Section
\ref{sec:sp4so4}, we will ignore the origin of the Higgs potential
 and adjust it by hand.

\subsection{Mooses and Low Energy Equivalence}
\label{sec:lowenergyequiv}

Mooses are a simple and flexible language to describe low energy
theories. Sites on a moose diagram correspond to the global/gauge
symmetries of the theory, and links correspond to fields that
transform as fundamentals or anti-fundamentals under the appropriate
symmetries.  In this paper, most link fields will be non-linear sigma
fields, so these descriptions will only be valid up to the scale of
Goldstone unitarity violation. 

The low energy equivalence is easy to understand in the moose language. 
The simplest moose relevant for electroweak physics is the standard
model with a Higgs boson.  Ignoring fermions, color, and hypercharge,
the electroweak sector is described by the following moose: 
\beq{higgsmoose}
\begin{tabular}{c}
\xymatrix@R=.5pc{\mathrm{Global:} & SU(2)_L && SU(2)_R \\
& *=<20pt>[o][F]{} \doublerightxyarrow^{\mbox{\raisebox{1.5ex}{$H$}}}
  && *=<20pt>[o][F]{} \\ \mathrm{Gauged:} &SU(2)_L && &} 
\end{tabular}
\ee
where $H$ can be written in terms of the usual Higgs doublet $h$ as
\be
H = \frac{1}{\sqrt{2}}\left(\begin{array}{cc}\epsilon\, h  &  h^\dagger \end{array}\right)= \frac{1}{\sqrt{2}}\left( \begin{array}{cc} h^0 & h^- \\ -h^+ & h^{0*}\end{array}\right), 
\ee
and $H$ transforms as $H \rightarrow g_L^\dagger H g_R$ under the $SU(2)_L \times SU(2)_R$ global symmetry.  The advantage of the $H$ notation over doublet notation is that \eq{higgsmoose} makes custodial $SU(2)$ symmetry manifest.

Now, it is straightforward to construct many different theories whose low energy physics is well described by \eq{higgsmoose}.  At energies much below the mass of the physical Higgs boson, we can simply replace $H$ with a non-linear sigma field $\Sigma$
\be
\Sigma = \frac{v_{\rm EW}}{2} e^{2 i \vec{\pi} \cdot \vec{\sigma} / v_{\rm EW}},
\ee
where $\vec{\sigma}$ are the Pauli matrices.  In the language of CCWZ
\cite{Coleman:1969sm,Callan:1969sn}, $\Sigma$ describes the Goldstone
bosons arising from the spontaneous breakdown of $SU(2)_L \times
SU(2)_R$ to the diagonal $SU(2)_V$.  Important for our purposes, there
are many ways to get a non-linear sigma model from a high energy
theory.  For example, in technicolor the $\Sigma$ field arises from a
fermion condensate 
\beq{technicolor}
\begin{tabular}{c}
\xymatrix@R=.5pc{\mathrm{Global:} & SU(2)_L && SU(2)_R & \\
& *=<20pt>[o][F]{} \rightxyarrow^{\mbox{\raisebox{1.5ex}{$\psi$}}} & *=<12pt>[o][F=]{} \rightxyarrow^{\mbox{\raisebox{1.5ex}{$\psi^c$}}} & *=<20pt>[o][F]{}  \\ \mathrm{Gauged:} &  SU(2)_L &SU(N_c)& }
\end{tabular}
\ee
where beneath $\Lambda_{TC}$, we can identify $\Sigma$ with fluctuations about the condensate $\langle \psi \psi^c \rangle$.  We can also generate a non-linear sigma model from a Wilson line in a flat or warped extra dimension.  Imposing the appropriate boundary conditions on an interval with a bulk $SU(2)$ gauge fields
\beq{extradimension}
\begin{tabular}{c}
\includegraphics[scale=0.6]{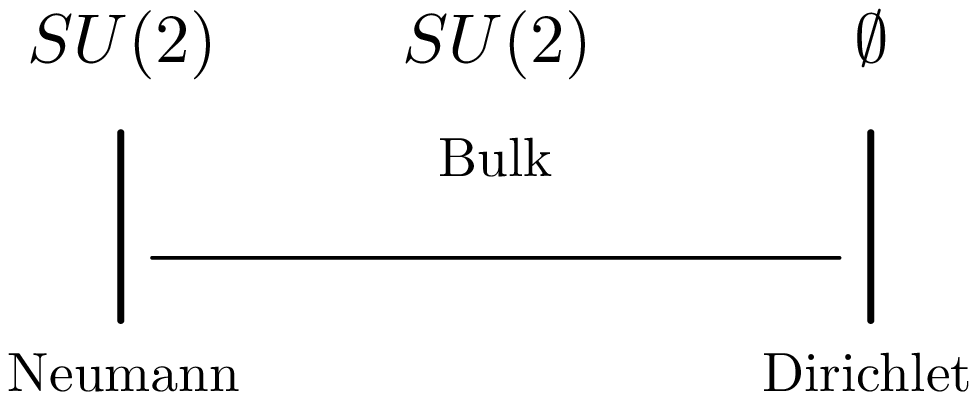}
\end{tabular}
\ee
the Wilson line $e^{i\int A_5 dx^5}$ has the same transformation
properties as $\Sigma$.  A particularly interesting extra dimensional
geometry is AdS$_5$, and \eq{extradimension} is expected to be dual to
a quasi-CFT with a gauged $SU(2)_L$ symmetry that is spontaneously
broken in the infrared \cite{Arkani-Hamed:2000ds,Rattazzi:2000hs},
\emph{i.e.}\ the Higgsless dual of technicolor. 

We can generate moose diagrams with additional sites with the same
light degrees of freedom by deconstructing these extra dimensions.
The geometry of the extra dimension is encoded in the different pion
decay constants on the various links
\cite{Cheng:2001nh,Randall:2002qr} 
\beq{Higgslessmoose}
\begin{tabular}{c}
\xymatrix@R=.5pc{\mathrm{Global:} & SU(2)_1 & SU(2)_2 & \qquad  & SU(2)_N & SU(2)_{N+1} \\
& *=<20pt>[o][F]{} \rightxyarrow^{\mbox{\raisebox{1.5ex}{$\Sigma_1$}}} &  *=<20pt>[o][F]{} \rightxyarrow^{\mbox{\raisebox{1.5ex}{$\Sigma_2$}}} &  *=<20pt>{\cdots} \rightxyarrow^{\mbox{\raisebox{1.5ex}{$\Sigma_{N-1}$}}}   &  *=<20pt>[o][F]{} \rightxyarrow^{\mbox{\raisebox{1.5ex}{$\Sigma_N$}}} & *=<20pt>[o][F]{} \\ \mathrm{Gauged:} &SU(2)_1 &SU(2)_2 &&  SU(2)_N && &}
\end{tabular}
\ee
The original $\Sigma$ field is given by
\be
\Sigma = \Sigma_1 \Sigma_2 \cdots \Sigma_N
\ee
and we can explicitly recover \eq{higgsmoose} from \eq{Higgslessmoose}
by integrating out sites corresponding to heavy gauge
bosons.\footnote{In general, integrating out sites from a moose will
  induce non-local interactions in theory space because the wave
  function of heavy gauge bosons span the entire space.  In the
  special case of AdS$_5$, these non-localities are suppressed because
  the heavy mode wave functions are localized \cite{Thaler:2005kr}.
  In any case, we can always capture the effect of theory space
  non-locality by introducing new interactions at higher order in the
  $\Sigma$ fields.}   

Finally, we can use the trick of hidden local symmetry
\cite{Bando:1987br} or little technicolor \cite{Thaler:2005kr} to
convert any non-linear sigma model into a moose diagram.  Using CCWZ,
any spontaneous symmetry breaking pattern can be described in terms of
a $G/H$ non-linear sigma model with a subgroup $F \subset G$ weakly
gauged.  We can then introduce a new ``$\rho$ meson'' gauge field to
generate the moose diagram 
\beq{ltmoose}
\begin{tabular}{c}
\xymatrix@R=.5pc{\mathrm{Global:} & G && G \\
& *=<20pt>[o][F]{}
  \doublerightxyarrow^{\mbox{\raisebox{1.5ex}{$\Sigma$}}} &&
  *=<20pt>[o][F]{} \\ \mathrm{Gauged:} &F && H&} 
\end{tabular}
\ee
The original CCWZ non-linear sigma model can be recovered from
\eq{ltmoose} by integrating out the $H$ gauge bosons.  Note that the
low energy moose from little technicolor is identical to the two-site
deconstruction of the warped AdS$_5$ dual theory
\cite{Contino:2003ve}.   

We have seen that a standard model Higgs, technicolor, extra
dimensions, quasi-conformal field theories, and general non-linear
sigma models all have descriptions in terms of mooses.  In the case of
electroweak physics circa 1980, the only relevant moose diagram for
discovering the $W$ and $Z$ bosons was 
\beq{1980moose}
\begin{tabular}{c}
\xymatrix@R=.5pc{\mathrm{Global:} & SU(2)_L && SU(2)_R \\
& *=<20pt>[o][F]{}
  \doublerightxyarrow^{\mbox{\raisebox{1.5ex}{$\Sigma$}}} &&
  *=<20pt>[o][F]{} \\ \mathrm{Gauged:} &SU(2)_L && U(1)_Y&} 
\end{tabular}
\ee
and until the precision electroweak tests from LEP experiments, in principle we
didn't even know whether the $W$ and $Z$ bosons were fundamental gauge
fields whose mass came from spontaneous symmetry breaking, or ``$\rho$
mesons'' from a strongly coupled theory as in the Abbott-Farhi model
\cite{Abbott:1981re,Abbott:1981yg}.  Even today, while precision
electroweak suggests a physical Higgs boson should exist in the form
of a linear sigma model UV completion for the standard model,
finely-tuned technicolor and Higgsless theories can still satisfy the
experimental bounds \cite{Cacciapaglia:2004rb,Foadi:2004ps}. 

The lesson from looking at different UV completions of the standard
model is that moose diagrams are a convenient way to organize one's
thinking about non-SUSY physics beyond the standard model.   Unlike
extra dimensional theories which yield a complete tower of KK modes,
one can adjust a moose deconstruction to only include modes relevant
for a given collider.   Moreover, mooses carry no implicit theoretical
biases and merely give a consistent framework to describe the relevant
spin-0 and spin-1 degrees of freedom.\footnote{There is no healthy
  lattice description of gravity, so while mooses can describe heavy
  spin-2 modes \cite{Arkani-Hamed:2002sp}, there is no straightforward
  extra-dimensional limit \cite{Arkani-Hamed:2003vb}.}    Of course,
the LHC will be able to tell the difference between a physical Higgs
and technicolor.  However, if we take the little hierarchy problem
seriously, then in the context of all current non-SUSY proposals, we
expect a non-linear sigma model description to suffice for at least
the initial running at the LHC, similar to the status of the standard
model in the pre-LEP era. 

\subsection{A Toy Little M-theory}
\label{sec:toy}

Little M-theories are classified according to their symmetry structure
and the embedding of the Higgs.  Because there are many different
symmetry breaking patterns that can yield a doublet charged under
$SU(2)_L \times U(1)_Y$ at low energies, there is no unique M-theory
to describe composite Higgs theories.  Rather, using the tools of the
previous subsection, every composite Higgs theory can be described by
a moose diagram, and in certain cases, one can interpolate between
different models by taking different limits of the same M-theory.  In
this subsection, we will show how this interpolation works in a toy
little M-theory without hypercharge or fermions.   

This toy model is based on the coset space $SU(3)/SU(2)$.  In particular, imagine a triplet of a global $SU(3)$ that takes a vacuum expectation value (vev).
\be
\Phi  = e^{i\Pi/f} \left(\begin{array}{c}0 \\0 \\f\end{array}\right)
\ee
The $SU(3)/SU(2)$ goldstone matrix contains a doublet $h$ and a singlet $\eta$ under the unbroken $SU(2)$.
\be
\Pi = \frac{1}{\sqrt{2}}\left(\begin{array}{ccc}0 & 0 & h_1 \\0 & 0 & h_2 \\h_1^\dagger & h_2^\dagger & 0\end{array}\right) + \frac{1}{2\sqrt{3}}\left(\begin{array}{ccc}\eta & 0 & 0 \\0 & \eta & 0 \\0 & 0 & -2\eta \end{array}\right)
\ee
There are at least three theories based on this coset space, namely the simple group little Higgs \cite{Kaplan:2003uc}, the minimal moose little Higgs \cite{Arkani-Hamed:2002qx}, and the original holographic Higgs \cite{Contino:2003ve}.  As we will see, they can all be described by the same three-site M-theory.  Further variations are discussed in \cite{Thaler:2005kr}. 

At first, it seems implausible that these three theories could arise as different limits of the same theory because they all have different fundamental gauge symmetries.  The minimal moose is based on gauging a product group $SU(2) \times SU(2)$, the simple group has the simple group $SU(3)$ gauged, whereas the original holographic Higgs is dual to a CFT with a single copy of $SU(2)$ gauged.  How can these theories come from the same M-theory if they have different gauge structures?  

The point is that for LHC phenomenology, we only require the low
energy degrees of freedom of the three theories to be the same, and
indeed, immediately above the electroweak symmetry breaking scale all
three theories have only massless $SU(2)$ gauge bosons.  The heavy
gauge fields will appear at the LHC as new heavy spin-1 modes, and in
the spirit of Abbott-Farhi, to first approximation we are free to
interpret these heavy modes as either gauge bosons that get a mass via
spontaneous symmetry breaking or resonances from some strong dynamics.
The little M-theory description will include an $SU(3) \times SU(2)$'s
worth of massive gauge bosons, but we can decouple any of the modes
that are irrelevant by changing some appropriate gauge couplings.   

The toy $SU(3)/SU(2)$ little M-theory can be described by the following moose diagram:
\beq{masterSU3moose}
\begin{tabular}{c}
\xymatrix@R=.5pc{\mathrm{Global:} & SU(3)_1 && SU(3)_m&& SU(3)_2 \\
& *=<20pt>[o][F]{} \doublerightxyarrow^{\mbox{\raisebox{1.5ex}{$\Sigma_1$}}} && *=<20pt>[o][F]{} \doublerightxyarrow^{\mbox{\raisebox{1.5ex}{$\Sigma_2$}}} && *=<20pt>[o][F]{} \\ \mathrm{Gauged:} &SU(2)_1 && SU(3)_m && SU(2)_2 &}
\end{tabular}
\ee
In unitary gauge, an $SU(3) \times SU(2)$'s worth of Goldstone are eaten, yielding $SU(3) \times SU(2)$ massive gauge bosons and massless $SU(2)$ gauge bosons.  The link fields are parametrized in terms of the uneaten Goldstones as
\be
\Sigma_1 = e^{i\Pi / f_1}, \qquad \Sigma_2 = e^{i \Pi / f_2}.
\ee
The $T$-parity limit of this theory is achieved when the gauge couplings $g_1$ and $g_2$ and the decay constants $f_1$ and $f_2$ are taken to be equal.  

It is now straightforward to see how \eq{masterSU3moose} can interpolate between the three different theories mentioned above.  If we take the $g_m$ gauge coupling to infinity, then we can integrate out the ultra-massive $SU(3)_m$ gauge bosons.  If we ignore the mechanism for generating the Higgs quartic, then this yields the correct gauge structure for the minimal moose: 
\be
\begin{tabular}{c}
\xymatrix@R=.5pc{\mathrm{Global:} & SU(3)_1 && SU(3)_2 \\
& *=<20pt>[o][F]{} \doublerightxyarrow^{\mbox{\raisebox{1.5ex}{$\Sigma$}}} && *=<20pt>[o][F]{} \\ \mathrm{Gauged:} &SU(2)_1 && SU(2)_2&}
\end{tabular}
\ee
where
\be
\Sigma = \Sigma_1 \Sigma_2.
\ee
The minimal moose exhibits a collective symmetry breaking structure, in that both $g_1$ and $g_2$ must be non-zero for the Higgs boson in $\Sigma$ to get a radiative potential from gauge loops.

If we take the $g_1$ and $g_2$ gauge couplings to infinity, then we can integrate out the ultra-massive $SU(2)_i$ gauge bosons.  This will yield the simple group little Higgs.  In order to see this, recall from \eq{ltmoose}, that in the little technicolor or hidden local symmetry construction, the moose
\be
\begin{tabular}{c}
\xymatrix@R=.5pc{\mathrm{Global:} & SU(3) && SU(3) \\
& *=<20pt>[o][F]{} \doublerightxyarrow && *=<20pt>[o][F]{} \\ \mathrm{Gauged:} & && SU(2)_\rho&}
\end{tabular}
\ee
turns into a $SU(3)/SU(2)$ nonlinear sigma model when the $SU(2)_\rho$ gauge boson is integrated out.  Therefore, when the $SU(2)_i$ gauge bosons are integrated out, we get a theory without an obvious moose description:
\be
\left(SU(3)/SU(2) \right)^2 \mbox{ non-linear $\sigma$-model with }  SU(3)_V \mbox{ gauged}
\ee
which is indeed the simple group theory.  Unlike the minimal moose, this theory does not exhibit ordinary collective symmetry breaking.  However, the Higgs potential is not quadratically divergent because both $f_1$ and $f_2$ must be nonzero for the Higgs boson not to be eaten.

Finally, \eq{masterSU3moose} can turn into the original holographic
Higgs if we take $f_1 > f_2$.  To see this, note that
\eq{masterSU3moose} can be thought of as the three-site deconstruction
of a warped extra dimension with bulk gauge fields and appropriate
boundary conditions: 
\be
\begin{tabular}{c}
\includegraphics[scale=0.6]{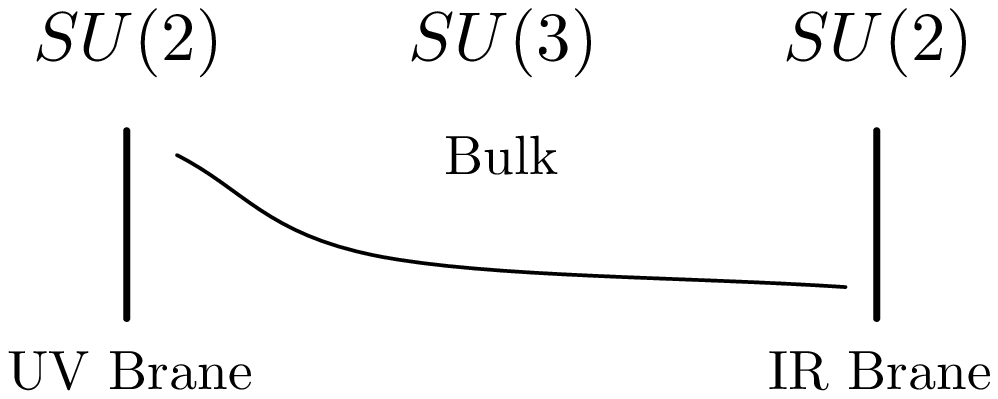}
\end{tabular}
\ee
The warp factor is reflected in the different pion decay constants on
the links, so there is no natural $T$-parity limit in this case.  The
original holographic Higgs exhibits AdS/CFT collective breaking, in
the sense that both the IR brane and UV brane boundary conditions must
violate the bulk $SU(3)$ symmetry in order for the Higgs to get a
radiative potential \cite{Thaler:2005en}.  To better reproduce an
extra dimension, we can add additional $SU(3)$ sites to the middle of
the moose. 

From a high energy perspective, these three theories have very
different philosophies, with different ``natural'' values for the
gauge couplings and the decay constants.  For the purposes of LHC
phenomenology, however, these theories are just models with novel
spin-1 and spin-0 spectra, and the M-theory description is a
convenient way to summarize their main features.   
In the next section, we describe a complete model with hypercharge,
custodial $SU(2)$, and three families of standard model fermions.

\section{The $Sp(4)/SO(4)$ Little M-theory}
\label{sec:sp4so4}

While the toy $SU(3)/SU(2)$ moose is good for illustrative purposes,
phenomenologically it is not the ideal model. The nonlinear sigma model
does not have a custodial $SU(2)$ symmetry which can lead to large
corrections to the $\rho$ parameter.  In addition, the up- and
down-type quark Yukawa couplings need to be implemented separately and
the down-type quark sector is not very appealing in the $SU(3)/SU(2)$
coset moose. In principle, it is also possible to write down a moose
model which can take a limit of the littlest Higgs. However, most of
the phenomenological studies of the little Higgs theories so far have
been focused on the littlest Higgs (see, \emph{e.g.}\
\cite{Han:2003wu}) It is therefore more useful to provide a model
which allows us to study collider phenomenologies for many alternative
theories in a uniform way.

The choice for this M-theory is not unique. One can always enlarge the
group structure and add more states to the theory so that it can
simulate more models and mimic them more accurately. However, it also
makes the theory unnecessarily complicated as one needs to break more
symmetries and to decouple many extra spurious states in considering
various limits. Therefore we will make a compromise between complexity
and versatility. We will choose a model which is simple enough to
describe and to be implemented in simulation tools and yet can still
capture a lot of interesting features of various models. One can
always extend the model to incorporate additional features when
necessary.  

We have made our choice based on the following goals. (1) We want a
theoretically consistent and calculable theory.  That is, the theory
should stay perturbative at energies below the cutoff and there should
be no gauge anomalies. (2) There should be no one-loop quadratic
sensitivity of the Higgs mass from the top Yukawa coupling and the
gauge couplings (except hypercharge), as this is main motivation for
much composite Higgs model building.  (3) The model should have a
custodial $SU(2)$ symmetry to protect $\rho$-parameter, and the Yukawa
structure should allow for minimal isospin breaking. (4) Minimal
flavor violation should be implemented to avoid large flavor-violating
effects. (5) To mimic models with extra dimensions we want
``KK-partners'' of all standard model fermions, but with a dial that
can decouple unnecessary particles. (6) The model should have a nice
$T$-symmetric limit because models with or without $T$-parity have
very different phenomenologies. (7) The model should have a lot of
dials which allow enough flexibitility to cover a variety of
phenomenolgy.

\subsection{Gauge/Higgs Sector}

We will choose the minimal symmetry structure with custodial $SU(2)$ that allows for both a simple group and a minimal moose limit. We need a group $G$ which contains an $SU(2)_L\times SU(2)_R =SO(4)$ subgroup and for which the coset space $G/SO(4)$ contains a Higgs doublet. The minimal choice is $G= SO(5) \simeq Sp(4)$. Note that $SO(5)/SO(4)$ contains a $\mathbf{4}$ of $SO(4)$, which looks like Higgs doublet with custodial symmetry. In our discussion we will use the language of $Sp(4)$ as the $SU(2)$ embedding is easier. Appendix A contains a summary of the $Sp(4)$ group for readers who are unfamiliar with the $Sp(2N)$ groups. As in the previous section, a three-site moose is chosen so that there is a nice geometric $T$-symmetric limit. 

The master $Sp(4)/SO(4)$ moose is as follows:
\beq{masterSp4moose}
\begin{tabular}{c}
\xymatrix@R=.5pc{\mathrm{Global:} & Sp(4)_1 && Sp(4)_m && Sp(4)_2 \\
& *=<20pt>[o][F]{} \doublerightxyarrow^{\mbox{\raisebox{1.5ex}{$\Sigma_1$}}} && *=<20pt>[o][F]{} \doublerightxyarrow^{\mbox{\raisebox{1.5ex}{$\Sigma_2$}}} && *=<20pt>[o][F]{} \\ \mathrm{Gauged:} &SU(2)_{L1} \times U(1)_R && Sp(4)_m && SU(2)_{L2} \times U(1)_R&}
\end{tabular}
\ee
Here we have chosen to gauge only one $U(1)$ with 
its generator given by 
\be
T_R = T_{3R1} + T_{3R2} + \frac{1}{2}(B-L).
\ee
After the $\Sigma$ fields take their vevs, the hypercharge generator is given by
\be
T_Y = T_R + T_{3Rm}.
\ee

There are several reasons for this choice of $U(1)$ charges, though
it is theoretically consistent to gauge two separate $U(1)$'s with generators given by
\beq{altu1}
T_{R1} = T_{3R1} + \frac{1}{4}(B-L), \qquad T_{R2} = T_{3R2} + \frac{1}{4}(B-L).
\ee
First of all, the quadratically divergent contribution induced by the $U(1)$ gauge interaction is not dangerous for cutoff $\sim$ 10 TeV due to the smallness of the gauge coupling. In fact, studies of the precision electroweak constraints on generic little Higgs theories~\cite{Csaki:2002qg,Hewett:2002px,Marandella:2005wd} show that the massive $U(1)$ gauge boson $A_H$ often causes the biggest problem so it is preferable to just gauge one $U(1)$ from that point of view. In the $T$-symmetric limit, on the other hand, even though there is no problem with electroweak constraints, it becomes very cumbersome to implement flavor with the extra $U(1)$.
A consequence is that the dark matter candidate $A_H$ in $T$-symmetric models is now replaced by the Goldstone mode that would have been eaten if we had gauged two $U(1)$s. This may affect the relic density calculation and the detection of the dark matter particles. It would make little difference for the collider phenomenology if it is difficult to tell the spin of the missing particles~\cite{Meade:2006dw}, assuming that either $A_H$ or the corresponding would-be eaten Goldstone boson is the lightest $T$-odd particle. 

The gauge bosons can be classified according to the generators they correspond to. There are $SU(2)_L$ gauge bosons on each site, $W_{1L}^{\pm,3},\, W_{2L}^{\pm,3}, W_{mL}^{\pm,3}$. For $SU(2)_R$ gauge bosons there is a complete set,  $W_{mR}^{\pm,3}$ in the middle site, but only one additional $W_R^3$ corresponding to $U(1)_R$. In addition, there are four more gauge bosons $X_m^{0,1,2,3}$ of the middle site corresponding to the off-diagonal generators of $Sp(4)$ ($T_{X(0,1,2,3)}$ as given in Appendix A). Their masses and mixings are shown in Appendix \ref{app:spectrum}.  Note that in this setup, there are no heavy $SU(3)_C$ gauge bosons.

Different models are reached by taking limits similar to the ones described in the previous section. Taking $g_m \to \infty$, we can integrate out the middle site and obtain the minimal moose model based on $Sp(4)$.  $T$-parity corresponds to taking $g_1=g_2$ and equal decay constants for $\Sigma_1$ and $\Sigma_2$. In fact, this moose is morally the same as the minimal moose model with $T$-parity~\cite{Cheng:2003ju} except that it has fewer links. We do not try to address the Higgs quartic potential through the little Higgs mechanism here because it is quite model dependent and it is not likely to be testable at the LHC other than finding a few more scalar states. On the other hand, if we take the gauge couplings $g_1$, $g_2$ of the $SU(2)_{1L}$ and $SU(2)_{2L}$ to infinity, we can integrate out the $SU(2)_{1L,2L}$ gauge bosons and obtain a simple group little Higgs based on the coset space $Sp(4)/SU(2)$ with extra $U(1)$s.  These limits are discussed further in Section \ref{sec:limits}.

The Higgs field is contained in
\be
\Sigma = \Sigma_1 \Sigma_2,
\ee
with  $\langle \Sigma \rangle = 1$.  It is convenient to choose a generator basis such that the $Sp(4)$ generators $T$ satisfy
\be
T A + A T^T = 0, \qquad A = \left(\begin{array}{cccc}0 & 1 & 0 & 0 \\-1 & 0 & 0 & 0 \\0 & 0 & 0 & -1 \\0 & 0 & 1 & 0\end{array}\right).
\ee
(See Appendix A). The $\Sigma$ vevs break the gauge symmetry $Sp(4) \times SU(2)^2 \times U(1)_R$ down to the standard model $SU(2)_L \times U(1)_Y$.  After the Goldstones are eaten, we are left with
\be
\Sigma = e^{i \Pi / f_{\rm eff}}, \qquad \Pi = \frac{1}{\sqrt{2}}\left(\begin{array}{cc} \begin{array}{cc}0 & 0 \\0 & 0\end{array} & H \\ H^\dagger & \begin{array}{cc}\phi^0/\sqrt{2} & \phi^- \\  \phi^+ & -\phi^0/\sqrt{2}\end{array} \end{array}\right)
\ee
where $H$ has the same definition as in Section~\ref{sec:lowenergyequiv}
\be
H = \frac{1}{\sqrt{2}}\left( \begin{array}{cc} h^0 & h^- \\ -h^+ & h^{0*}\end{array}\right),
\ee
and 
\be
\frac{1}{f_{\rm eff}^{2}}=\frac{1}{ f_1^{2}}+\frac{1}{f_2^{2}}.
\ee
We see that there are three extra Goldstone bosons in addition to the
Higgs. They would have been eaten if we had chosen to gauge a whole
$SU(2)_R$ on one of the boundary sites.  Indeed, this is precisely
what happens in the minimal holographic Higgs model
\cite{Agashe:2004rs}. However, this choice for the gauge structure
would prevent us from taking the $T$-symmetric limit.\footnote{It is
  in principle plausible to add a switch to change between these cases
  when implementing the model into a collider simulation program.
  Note that if one were to make this choice, then the fermion sector
  described in the next subsection would have to be modified.}   

The mechanisms for generating the Higgs potential in little Higgs
theories are quite model-dependent and there can be one or two (or
even more) light Higgs doublets. In the simplest little Higgs
model~\cite{Schmaltz:2004de}, the Higgs potential is generated through
the top loop and requires some mild fine-tuning. The minimal moose
model contains many link fields and the Higgs potential is given by a
collection of complicated plaquette operators. Given this
model-dependence and the fact that the LHC is unlikely to test the
little Higgs structure of the Higgs self-couplings, we will simply
write down the necessary Higgs potential for the electroweak symmetry
breaking without specifying its origin.  We show how to do this in a
theoretically consistent way in Appendix
\ref{sec:goldstonemasses}. This is probably good enough if there is
only one light Higgs within the reach of the LHC.  It is possible to
extend the symmetry structure or number of link fields to allow for
multiple Higgses if necessary.  

\subsection{Fermion Sector}

We will introduce fermions on each site and the standard model
fermions will come from various linear combinations. By changing the
linking masses we can change the profiles of the ``zero mode''
fermions which then looks like localization in extra dimensions. This
freedom also allows us to localize the standard model fermions away
from the site(s) where the gauge coupling is taken to be large in
various limits. On the other hand, it also implies that there will be
excited (``KK'') fermions. To avoid large flavor-changing effects, we
would like to implement minimal flavor violation which either means
that the heavy fermions should share the Yukawa structure of the
standard model fermions or that they are degenerate among
generations. Choosing to consider the $T$-parity limit forces a small
amount of model building upon us, and some additional fields will be
needed so that we can independently control the masses of the $T$-odd
fermions. 

If we were only interested in third generation quarks, then the
simplest fermion sector to get the top and bottom Yukawa couplings
would look like 
\beq{minimalfermion}
\begin{tabular}{c}
\xymatrix@R=.5pc{\mathrm{Gauged:} &SU(2)_{L1} \times U(1)_R && Sp(4)_m
  && SU(2)_{L2} \times U(1)_R& \\ 
& *=<20pt>[o][F]{}
  \doublerightxyarrow^{\mbox{\raisebox{1.5ex}{$\Sigma_1$}}} &&
  *=<20pt>[o][F]{}
  \doublerightxyarrow^{\mbox{\raisebox{1.5ex}{$\Sigma_2$}}} &&
  *=<20pt>[o][F]{} \\ \mathrm{Quarks:} &  Q_1^c && Q_m && Q_2^c} 
\end{tabular}
\ee
where
\be
Q_i^c = \left(\begin{array}{c}0 \\t_i^c  \\ b_i^c \end{array}\right), \qquad 
Q_m = \left(\begin{array}{c}q_m \\ t_m \\ b_m  \end{array}\right).
\ee
From the only interactions local in theory space
\beq{minimalyukawa}
-\mathcal{L}_{\rm Yukawa} = f_1 Q_m^T \Sigma_1^\dagger \beta_1 Q_1^c +
f_2 Q_m^T \Sigma_2 \beta_2 Q_2^c + \mathrm{h.c.}, 
\ee
we would generate top and bottom Yukawa couplings and masses for heavy
top and bottom partners. The splitting between the top and the bottom
comes from the fact that $\beta_i$, $i=1,\ 2$, are actually custodial
$SU(2)$-violating matrices, $\beta_i={\rm
  diag}(\beta_{qi},\beta_{qi},\beta_{ti},\beta_{bi})$.   

There are two reasons why we want to expand this minimal setup.
First, \eq{minimalfermion} only has $SU(2)_L$ singlet fermion
partners.  More generally, we expect there to be regions of model
space with $SU(2)_L$ doublet fermion partners, so we would like to
augment \eq{minimalfermion} with additional heavy doublet states that
may or may not be decoupled.  Note that because of the gauge
symmetries and our demand for theory space locality, we cannot have
just $SU(2)_L$ doublet fermion partners; if we were to choose $Q_i^c =
( q_i^c \; 0 )$ (and reverse the $SU(3)_C$ charges of all the fields),
there would be no way to split the top and bottom Yukawa couplings
using the interaction in \eq{minimalyukawa}. 

Second, we want to treat all three fermion generations symmetrically
in order to implement minimal flavor violation. However, in the
$T$-symmetric limit, \eq{minimalyukawa} tells us that there is a fixed
ratio between the SM fermion  and their $T$-odd partners masses,
i.e. $m_{q'} \sim (m_q/v_{\rm EW})f_{\rm eff}$.
Therefore, for any reasonable ratio of $v_{\rm EW}$ to $f_{\rm eff}$,
the $T$-odd partner of, say, the electron would be much lighter than
the $Z$ and therefore excluded by the LEP bounds on the $Z$ width.  In
order to treat the three generations symmetrically, we need a
mechanism such that the $T$-odd partners will all be at the TeV scale,
but there will still be a hierarchy among the $T$-even standard model
fields.  In order to do so, we will assume that the $\beta_i$'s are
nearly degenerate for all three generations, then will we use a
see-saw mechanism to decrease the effective Yukawa coupling for the
lighter fermions.  While this may see like excessive model building
for a phenomenological model, we emphasize that our goal is to have a
description of non-SUSY LHC physics that is aware of the various model
building challenges that the little hierarchy problem presents.

Visually, for each generation, the complete fermion sector is
\be
\begin{tabular}{c}
\xymatrix@R=.5pc{\mathrm{Gauged:} &SU(2)_{L1} \times U(1)_R && Sp(4)_m && SU(2)_{L2} \times U(1)_R& \\
& *=<20pt>[o][F]{}
  \doublerightxyarrow^{\mbox{\raisebox{1.5ex}{$\Sigma_1$}}} &&
  *=<20pt>[o][F]{}
  \doublerightxyarrow^{\mbox{\raisebox{1.5ex}{$\Sigma_2$}}} &&
  *=<20pt>[o][F]{} \\ \mathrm{Quarks:} & Q_1, Q_1^c && Q_m && Q_2,
  Q_2^c \\ 
\mathrm{Leptons:} & L_1, L_1^c && L_m && L_2, L_2^c}
\end{tabular}
\ee
with floating fermions $Q', Q'^c, L', L'^c$ to enable the flavor 
see-saw mechanism.\footnote{The floating fermions violate theory space locality, and along with the non-local definition of $U(1)_R$, prevent us from taking a strict holographic composite Higgs limit.  On the other hand, without the floating fermions, we know of no way to implement minimal flavor violation in the $T$-symmetric limit.  This tension between $T$-parity, flavor structure, and theory space locality has been observed before \cite{Thaler:2005en}, and is probably a generic issue for composite Higgs theories.}   Using the third generation as an example, the
fermions are embedded as 
\be
Q_i = \left(\begin{array}{c}q_i \\0 \\ 0 \end{array}\right), \quad
Q_i^c = \left(\begin{array}{c}q_i^c \\t_i^c  \\ b_i^c
\end{array}\right), \quad  
Q_m = \left(\begin{array}{c}q_m \\ t_m \\ b_m  \end{array}\right), \quad 
Q'= \left(\begin{array}{c}0 \\ t' \\ b' \end{array}\right), \quad 
Q'^c = \left(\begin{array}{c}0 \\ t'^c \\ b'^c \end{array}\right),
\ee
\be
L_i = \left(\begin{array}{c}\ell_i \\0 \\ 0 \end{array}\right), \quad
L_i^c = \left(\begin{array}{c}\ell_i^c \\ \nu_i^c  \\ \tau_i^c
\end{array}\right), \quad 
L_m = \left(\begin{array}{c}\ell_m \\ \nu_m \\ \tau_m
\end{array}\right), \quad  
L'= \left(\begin{array}{c}0 \\ \nu' \\ \tau' \end{array}\right), \quad 
L'^c = \left(\begin{array}{c}0 \\ \nu'^c \\ \tau'^c \end{array}\right).
\ee
The anomaly-free fermion charges are given in Figure
\ref{fig:fermioncharges}.   Note that $Q', Q'^c, L', L'^c$ contain
only the lower two components which are only charged under $U(1)_R$,
so they do not need to be associated with any site. 

\FIGURE{
\parbox{6in}{$$
\renewcommand\arraystretch{1.5}
\begin{array}{c||cccccc|cccccc}
                          &  & & & (\overline{q_i})
             &(t'^c,\overline{t'}) & (b'^c,\overline{b'})&  &  &  &
             (\overline{\ell_i})  & (\nu'^c,\overline{\nu'}) &
             (\tau'^c, \overline{\tau'}) \\  
             &  q_m & t_m & b_m & q_i^c & t_i^c & b_i^c &  \ell_m & \nu_m & \tau_m & \ell_i^c & \nu_i^c & \tau_i^c \\ 
       \hline
       \hline
SU(3)_C        &  \mathbf{{3}} & \mathbf{{3}} & \mathbf{{3}} & \mathbf{\bar{3}}& \mathbf{\bar{3}}& \mathbf{\bar{3}} & \mbox{--}& \mbox{--}& \mbox{--}& \mbox{--}& \mbox{--}& \mbox{--}\\
\hline
SU(2)_{Lm} &  \mathbf{2} & \mbox{--}& \mbox{--}& \mbox{--}& \mbox{--}& \mbox{--} &  \mathbf{2} & \mbox{--}& \mbox{--}& \mbox{--}& \mbox{--}& \mbox{--}\\
\hline
SU(2)_{Li}   &  \mbox{--} & \mbox{--} & \mbox{--} & \mathbf{2} & \mbox{--} & \mbox{--} & \mbox{--}&  \mbox{--} &  \mbox{--} & \mathbf{2}  & \mbox{--} & \mbox{--}  \\
\hline
U(1)_{3Rm}  &  0 & \frac{1}{2} & - \frac{1}{2} & 0& 0&  0 &  0 & \frac{1}{2} & - \frac{1}{2} & 0& 0& 0\\
\hline
U(1)_{R}     &  \frac{1}{6} & \frac{1}{6} & \frac{1}{6}& -\frac{1}{6} & -\frac{2}{3}  & \frac{1}{3} &  -\frac{1}{2} & -\frac{1}{2} & -\frac{1}{2}& \frac{1}{2}  & 0 & 1 \\
\hline
\end{array}
\renewcommand\arraystretch{1.0}
$$}
\caption{The anomaly-free fermion charges for the $Sp(4)/SO(4)$ moose.  Note that we have decomposed $Sp(4)_m$ as $SU(2)_{Lm} \times U(1)_{3Rm}$.}
\label{fig:fermioncharges}
}
We will now write down all couplings local in theory space as well as
the leading non-local interaction.  Each of these couplings preserves
enough of an $Sp(4)$ global symmetry to avoid one-loop quadratically
divergent contributions to the Higgs potential.  The reason for
including the leading non-local interaction is that in the minimal
moose-like limit, the center site is strongly coupled so we want
standard model fields to live on the outside sites in order to avoid
large four-fermion operators.  The non-local interaction will then
provide the dominant Yukawa couplings in this limit (See Section 4 for
a more detailed discussion). 

The Yukawa interactions and masses of the fermions are given by

\be
-\mathcal{L}_{\rm Yukawa} = f_1 Q_m^T \Sigma_1^\dagger \beta_1 Q_1^c + f_2 Q_m^T \Sigma_2 \beta_2 Q_2^c + \mathrm{lepton~terms} + \mathrm{h.c.},
\ee
\be
-\mathcal{L}_{\rm Mass} = m_{1} Q_1^T Q_1^c + m_{2} Q_2^T Q_2^c + \mathrm{lepton~terms} +\mathrm{h.c.}.
\ee
\be
-\mathcal{L}_{\rm Non-Local} = \alpha_1 f_{\rm eff} Q_1^T \Sigma_1 \Sigma_2 Q_2^c + \alpha_2 f_{\rm eff} Q_2^T \Sigma_2^\dagger \Sigma_1^\dagger Q_1^c + \mathrm{lepton~terms} +\mathrm{h.c.},
\ee
where $\beta_i$'s are coupling matrices mentioned before.
To get small Yukawa couplings to produce the observed fermion mass hierarchies, we can use a see-saw mechanism with the help of $Q', Q'^c$
\be
-\mathcal{L}_{\rm See-Saw} = {Q'}^T (F_1 Q_1^c - F_2 Q_2^c) + {Q'}^T  K Q'^c + \mathrm{lepton~terms}+ \mathrm{h.c.},
\ee
where $F_i={\rm diag}(0,0,F_{Ti}, F_{Bi})$ and $K= {\rm diag}(0,0, K_{T}, K_{B})$. In the limit that $F_{Ti,Bi} \gg \beta v_{\rm EW}, K_{T,B}$, the effective Yukawa couplings for the standard model fermions are suppressed by 
\be
\lambda_{\rm eff} \propto \frac{K}{F}.
\ee
For three generations of fermions, $F_{Ti},\, F_{Bi},\, K_T,\,  K_B$
are $3\times 3$ matrices which encode the flavor structure.  As we discuss in the next section, minimal flavor violation corresponds to allowing all of the Yukawa structure to appear only in the $K$ matrix.

\section{Exploring the Parameter Space of Little M-theory}
\label{sec:limits}

Just like in the MSSM, many parameters in little-M theory are
already constrained by experimental data.  In this section, we will give a general discussion of
the constraints one should be aware of when exploring the little M-theory parameter space.  This will also give us an opportunity to show how to approach various ``natural'', or more familiar, limits.

Unfortunately, because the $U(1)_R$ gauge coupling violates theory
space locality, strictly speaking we cannot take a holographic composite Higgs limit as we did in Section \ref{sec:toy}.  Then again, if we used the $U(1)$ charge assignments of \eq{altu1}, then the model would have an
extra-dimensional limit as long as the floating $Q'$ and $L'$ fermions
were decoupled (\emph{i.e.}\ if the $K$ parameter were taken to
infinity).  Notice that the holographic limit also does not
simultaneously allow for $T$-parity. It would be interesting to
explore the phenomenology along that direction, but we
will not explicitly discuss this limit further here.

Ignoring the Goldstone sector, the parameter space of little M-theory is defined by
\bea
\mbox{Decay Constants:} && f_1, \quad f_2, \nonumber\\
\mbox{Gauge Couplings:} && g_1, \quad g_m, \quad g_2, \quad g_R, \nonumber\\
\mbox{Fermion Parameters:} && \beta_i, \quad m_i, \quad \alpha_i, \quad F_i, \quad K.
\eea
Roughly speaking, the parameter space is spanned by two orthogonal directions: 1) whether $T$-parity is a good symmetry or not, and 2) whether the model is more simple-group-like or more product-group like.  At first glance, both of these directions seem to mainly affect the values of the gauge parameters.  Indeed, the $T$-parity axis corresponds to splitting $g_1$ and $g_2$, and the gauge structure axis corresponds to either taking $g_m$ or $g_{1,2}$ large.  

As we will see, however, the fermion parameters should be adjusted in concert with the gauge parameters in order to satisfy various experimental bounds.  Four-fermion operators generated by integrating out heavy ($T$-even in the $T$-parity limit) gauge bosons generically present the
strongest constraints on the little M-theory parameter space.   In general, one should choose
the fermion parameters in such a way that  the standard model fermions and gauge bosons have approximately the same profiles in theory space, so that the overlaps between the standard model
fermions and the heavy gauge bosons are small.  

Regardless of the values of the gauge parameters, flavor- and isospin-violating effects will always impose constraints on the fermion parameters, so we will discuss those bounds first.  If we choose to impose minimal flavor and isospin violation, the dominant constraints on the values of the little M-theory parameters come from four-fermion operators involving standard model fields and electroweak precision tests.  We will discuss those bounds by first exploring the $T$-symmetric limit and then seeing how the model building constraints change at the simple group and minimal moose limits on the gauge structure axis.

\subsection{Minimal Flavor and Isospin Violation}

Because the heavy ``KK partners'' of the standard model fermions
obtain their masses from $F_i$, $\beta_i f_i$, $m_i$, and $\alpha_i
f_{\rm eff}$, flavor-changing effects will be induced if these
parameters are not flavor universal.  To avoid large flavor-changing
effects from the heavy fermions it is simplest to implement minimal
flavor violation.  That is, we can take all the above parameters to be
flavor universal and put all the flavor structure in the $K$ matrices,
such that $K_{T,B}$ are proportional to the observed Yukawa matrices
of the standard model fermions.   If we ignore inter-generational
mixings, the fermion mass eigenstates for each species are obtained by
diagonalizing a $5\times 5$ mass matrix and there are 2 ``KK'' modes
for each handedness. Some of them may be decoupled in various
limits. The mass matrices for the fermions are discussed in Appendix
\ref{app:spectrum}. 

The various terms in the fermion Lagrangian also induce mixings
between $SU(2)_L$ doublet fermions and singlet fermions, which can
result in isospin-violating shifts of the standard model fermion
couplings to the $W$ and $Z$ bosons of the order
$v^2/(f^2,F^2,m^2)$. LEP and SLC experiments have tested these
couplings of the light fermions to a precision level of $10^{-3}$
\cite{isospinviolation}. The simplest solution to this constraint is
to have these parameters (except $K$) respect the isospin symmetry,
such that $\beta_{ti}=\beta_{bi}$, $F_{Ti}=F_{Bi}$, and so on for (at
least) the light generation quarks and leptons.  It is possible to
deviate from the assumption of minimal flavor and isospin violation in
various corners of the parameter space (especially for the third
generation), but one then needs to check various experimental
constraints such as $Z\to b\bar{b}$ case by case. 

\subsection{The $T$-symmetric Limit}

The motivation for considering a $T$-symmetric limit is that $T$-odd
particles cannot lead to tree-level modifications of precision
electroweak measurements or four-fermion operators.  In the $T$-symmetric limit, the
number of free parameters are greatly reduced. We have $g_1=g_2\equiv
g,\, f_1=f_2\equiv f= \sqrt{2} f_{\rm eff},\, m_1=m_2\equiv m,\,
\beta_1=\beta_2\equiv \beta, \, \alpha_1=\alpha_2\equiv \alpha,\,
F_1=F_2\equiv F$. The mass eigenstates divide into $T$-even and
$T$-odd states. $T$-parity is defined as the geometric symmetry of the
moose diagram by exchanging site-1 and site-2, with a twist by
$\Omega={\rm diag}(1,1,-1,-1)$. The $\Omega$ twist flips the parity of
the bottom two components of the fermions and the off-diagonal
$2\times2$ blocks of the gauge fields and the Goldstone fields, so
that all standard model fields (including the Higgs) are even under
$T$-parity.   

The even and odd states in the gauge sector are
\bea
\mbox{$T$-even:}&& \quad W_{+L}^{\pm,3} \equiv
\frac{1}{\sqrt{2}}(W_{1L}^{\pm,3}+W_{2L}^{\pm,3}), \;\;
W_{mL}^{\pm,3},\;\; W_R^3,\;\; W_{mR}^{\pm,3} \nonumber \\ 
\mbox{$T$-odd:}&& \quad W_{-L}^{\pm,3} \equiv
\frac{1}{\sqrt{2}}(W_{1L}^{\pm,3}-W_{2L}^{\pm,3}), \;\; X^{\pm} \equiv
\frac{1}{\sqrt{2}}(X^2 \mp i X^1), \;\; \nonumber \\
&& \quad X^{n(*)} \equiv \frac{1}{\sqrt{2}} ( X^0 \pm iX^3) 
\eea
and in the fermion sector are
\bea
\mbox{$T$-even:}&& \quad q_m,\;\; q_+
\equiv\frac{1}{\sqrt{2}}(q_1+q_2),\;\; q_+^c \equiv
\frac{1}{\sqrt{2}}(q_1^c+q_2^c),\;\;  \nonumber\\
&&  \quad t_+^c (b_+^c) \equiv
\frac{1}{\sqrt{2}}\Big( t_1^c(b_1^c)- t_2^c(b_2^c)\Big), \; \; t'(b'),\;\; t'^c(b'^c) \nonumber
\eea
\bea 
\mbox{$T$-odd:}&& \quad t_m(b_m),\;\; q_-\equiv\frac{1}{\sqrt{2}}(q_1-q_2),\;\; q_-^c \equiv
\frac{1}{\sqrt{2}}(q_1^c-q_2^c),\;\; \nonumber\\
&& \quad t_-^c (b_-^c) \equiv
\frac{1}{\sqrt{2}}\Big( t_1^c(b_1^c)+ t_2^c(b_2^c)\Big), 
\eea
and in the scalar sector are
\bea
\mbox{$T$-even:}&& \quad H, \nonumber \\
\mbox{$T$-odd:}&& \quad \phi^{\pm,0}.
\eea
States with same quantum numbers under the standard model and same
$T$-parity mix in general. Of course after electroweak symmetry
breaking, there are also mixings between states transforming under
$SU(2)_L$ and states transforming under $SU(2)_R$.  The masses of
these states are calculated in Appendix \ref{app:spectrum}.  

In the $T$-symmetric limit with minimal flavor and isospin violation,
the dominant constraints on the values of the little M-theory
parameters come from the four-fermion operators involving standard
model fields and loop corrections to the $Z\to b \bar{b}$ vertex and
the $\rho$ parameter.  At tree-level, the four fermion operators come
from integrating out heavy $T$-even gauge bosons, and we will discuss
these constraints more in the next subsection.  The loop corrections
only impose mild constraints and the $f$'s can be well below 1
TeV~\cite{Cheng:2004yc}, making the new particles more accessible at
the LHC.  In fact, the $T$-symmetric limit is probably the least
constrained scenario, and we expect there to exist a large region of
``safe'' parameter space both at and near the $T$-parity limit.  Of
course if we deviate from minimal flavor and isospin violation, there
are other constraints to worry about but they are model-dependent.  

Deviations from $T$-parity are generically subject to stronger
electroweak precision constraints, which would have to be checked by
hand.  Without $T$-parity, one can have tree-level modifications of
precision electroweak parameters.  Having said that, we expect that
for the $Sp(4)/SO(4)$ model with a custodial $SU(2)$ symmetry and only
one $U(1)$ gauged, the constraints from oblique parameters should be
weak.  We do expect four-fermion operators to be dangerous away from
the $T$-parity limit. 

\subsection{The Gauge Structure Axis}

As we adjust the ratio of $g_m$ to $g_{1,2}$, we interpolate between simple group and product group gauge structures at low energies.  However, if we are not careful, then the effect of the ``decoupled'' heavy gauge bosons can be large.  The mass of the heavy gauge bosons scale as $\sim g f$ but their couplings scales like $~\sim g$, so at low energies one generically once expects to find non-decoupling four-fermion operators suppressed only by $1/f_{\rm eff}^2$.  In order to soften the effects of these heavy gauge bosons, we should arrange the standard model fermion wavefunctions in theory space to have small overlaps with the strong-coupling site(s).  

In the next subsection, we show explicitly how to minimize the size of four-fermion operators at the two extremes of the gauge structure axis, and comment further on precision electroweak constraints away from the $T$-parity limit.  In both the simple group and minimal moose limits, it is also possible to decouple unnecessary modes if one does not wish to impose $T$-parity.

\subsubsection{The Simple Group Limit}

As already mentioned, the simple group little Higgs limit is obtained
by taking the gauge couplings $g_1$ and $g_2$ of the outer sites to
infinity. In this case, the $SU(2)_L$ gauge bosons on sites 1 and 2,
$W_{1L}^{\pm,3}$ and $W_{2L}^{\pm,3}$, become heavy and decouple from
the low energy spectrum. To further decouple their effects in order to
avoid possible large four-fermion interactions induced by them
\cite{Low:2004xc}, the standard model $SU(2)$-doublet fermions should
be mostly localized in the middle site. This can be done by taking
$m_1$ and $m_2$ large. In the limit $m_{1,2}\to \infty$, $q_1$,
$q_1^c$, $q_2$, $q_2^c$ decouple and the fermion spectrum
simplifies. If one does not need to take the $T$-symmetric limit, the
Yukawa flavor structure can come from one of the $\beta$'s. In this
case the fermion sector can be further simplified by taking $K$ to
infinity to decouple $t'(b')$, $t'^c(b'^c)$. 
The standard model Yukawa couplings can be obtained by taking
\be
\beta_{t(b)1} \ll \beta_{t(b)2} \sim \mathcal{O}(1),
\ee
and $t_m(b_m)$ and $t_2^c(b_2^c)$ will acquire a mass of
$\beta_{t(b)2}f_2 \sim {\rm TeV}$. The $\beta_{t(b)2}$ should be
family-universal to avoid large flavor-changing effects. The standard
model fermions and Yukawa couplings are approximately given by $q_{\rm
  SM} \sim q_m$, $t^c_{\rm SM} (b^c_{\rm SM}) \sim t_1^c(b_1^c)$ and
$\lambda_{t(b)} \sim \beta_{t(b)1}\,s/2$, where $s=
f_2/\sqrt{f_1^2+f_2^2}$, up to small corrections. 

In the $SU(3)/SU(2)$ simple group little Higgs model, the strongest
constraints come from the $Z-Z'$ mixing and the four-fermion interactions
induced by $Z'$~\cite{Schmaltz:2004de}, where $Z'$ is the gauge boson
corresponding to the generator $T_8$. As a result $\sqrt{f_1^2+f_2^2}$
is required to be larger than a couple TeV. The fine-tuning can be
reduced by taking unequal $f_1$ and $f_2$ which allows the freedom to
adjust the masses of the top partner and gauge partners relatively. In
our model, on the other hand, there is a complete custodial $SU(2)$
multiplet of $W_{mR}^{\pm,3}$ gauge bosons which can mix with the
standard model $W_{mL}^{\pm,3}$ gauge bosons after electroweak
symmetry breaking. As a result, the mixing does not induce any further
custodial $SU(2)$ violation unlike in the $SU(3)/SU(2)$ model. In
addition, the standard model fermions are not charged under the
$SU(2)_{Rm}$ subgroup of $Sp(4)_m$. The only four-fermion operators
induced by heavy gauge bosons at the leading order (not suppressed by
$v^2/f^2$) come from the $U(1)_R$ component of $Z_R$, which is a
combination of $W_{mR}^3$ and $U(1)_R$ gauge field $W_R^3$ (see
Appendix B). However, they are suppressed by the smallness of the
component $W_R^3$ in $Z_R$ and the $U(1)_R$ gauge coupling. The
constraint from the electroweak precision measurements can therefore
be much weaker. 

\subsubsection{The Minimal Moose Limit}

The minimal moose little Higgs limit is obtained by taking $g_m$ to
infinity. In this case we can integrate out the middle site and all
gauge bosons of the middle site decouple. To avoid large residual
four-fermion interactions induced by them, the standard model fermions
should live away from the middle site. This can be achieved by taking
$\beta_{q1} f_1$, $\beta_{q2} f_2$ to be larger than $\alpha_2 f_{\rm
  eff}$, $m_1$, $\alpha_1 f_{\rm eff}$, $m_2$. More specifically, if
we do not need to take the $T$-symmetric limit, we can remove $Q_m$
and $Q_2^c$ from the low energy spectrum as follows: take $\beta_{q2}
f_2$ large ($\gg m_2, \alpha_1 f_{\rm eff}$) so that $q_m$ and $q_2^c$
decouple, and $\beta_{t(b)2}f_2$ large ($>\beta_{t(b)1} f_1$) so that
$t_m(b_m)$ and $t_2^c(b_2^c)$ decouple. We are left with a complete
$Sp(4)$ multiplet $Q_1^c= (q_1^c, t_1^c, b_1^c)^T$ and $q_1$, $q_2$,
$t'(b')$, $t'^c(b'^c)$. Two pairs of the quark-antiquark get $\sim$
TeV masses from $m_1$ and $F_{T(B)1}$. They play the roles of the
heavy fermions which cancel the one-loop quadratic divergence from the
standard model top quark loop. The Yukawa couplings are
$\lambda_{t(b)} \sim \alpha_2 K_{T(B)}/ (2 F_{T(B)1})$ (assuming
$\alpha_2 f_{\rm eff} \ll m_1$). With minimal flavor and isospin violation, the
standard model Yukawa structure and the top-bottom splitting all come
from $K_{T,B}$. The above discussion is more transparent by examining
the fermion mass matrix given in Appendix~B. 

Because the model has a custodial $SU(2)$ symmetry and we only gauge
one $U(1)$, the strongest constraint comes from the remaining four-fermion
interactions generated by the $W'_L$ and $Z'_L$ gauge bosons (the
heavy combinations of the $W_{1L},\, W_{2L}$ and the $U(1)_R$ gauge
bosons). They are required to be heavier than a few TeV if the
standard model fermions are localized on one site, but the constraint
can be greatly relaxed by taking the $T$-symmetric limit.  In particular, the $W'_L$ and
$Z'_L$ gauge bosons are $T$-odd, therefore they cannot contribute at
tree-level to any four-fermion or precision electroweak operator. 

\section{Comments on Collider Phenomenology}

Little M-theory is a framework which captures the dominant features of
composite Higgs models. We expect this framework to exhibit a rich
phenomenology with many novel features which deserve detailed
studies.  In particular, a   
comprehensive study of the inverse map from LHC signature space to
little M-theory parameter space could and should be undertaken, and it
will be interesting to see whether little M-theory with $T$-parity
exhibits the same degeneracy structure as was found in the
MSSM~\cite{Arkani-Hamed:2005px}.  In this section, we confine
ourselves to qualitative comments about the phenomenology of little
M-theory.   
The collider phenomenologies are very different for models with or
without $T$-parity. 
We start our discussion in the $T$-symmetric limit because it is
preferred by experimental constraints.

The initial signal of a $T$-symmetric composite Higgs scenario will be significant and
dramatic, because the QCD pair production cross-section for the fermionic partners of standard model quarks will be large. The decays of the $T$-odd quark partners typically proceed through decay chains
involving gauge boson partners as well as leptonic partners, terminating in the lightest $T$-odd particle. Therefore, the typical signature will be jets, leptons and
large missing energy. Without measuring at least some details of the
spectrum and couplings, it is probably indistinguishable from a
supersymmetric scenario. Therefore, it should be included, along with
SUSY, in the collection of initial candidates of possible scenarios if such
signatures are found. 

There will be exotic gauge bosons, like $W_R^{\pm},\, Z_R$ ($T$-even) and
$X$ ($T$-odd). ($Z_R$ is a massive combination of $W_{mR}^3$ and the
$U(1)_R$ gauge field $W_R^3$ defined in Appendix B.)  Generically, it
is not possible to make all of the heavy 
gauge bosons on the middle site odd under $T$-parity. This is an
interesting difference 
between a pure product group structure, which only requires the two
outside sites, and a simple group structure, which is represented here
by the middle site. Because of this fact, a smoking gun signature for
the existence of some simple group structure is the existence of these
exotic gauge bosons.  
In this model, $Z_R$ can easily be seen as a resonance by Drell-Yan production.
However, this is not a distinguishing feature between little-M theory
and SUSY, as 
extensions of the MSSM could certainly have new gauge sectors.
$W_R^\pm$ only couples to standard model fermions through mixings after
electroweak symmetry 
breaking. The couplings are suppressed by ${\cal O}(v^2/f^2)\sim
10^{-2}$, so they will be difficult to produce directly. $X$ gauge
bosons are $T$-odd so they need to be pair-produced or produced
together with another $T$-odd state. Their decays will have missing
energy which makes it a challenge to reconstruct their identities. 
In the case of $g_1 \sim g_2 \sim g_m$, we will be able to produce
both heavier combinations of $SU(2)_L$, $W'_L$, $W_L^{\prime \prime}$ (from
\eq{eq:wmassmatrix}), as well as some of the exotics. It is obviously an 
interesting new benchmark to explore. In particular, the verification
of cancellation of quadratic divergences is expected be more
complicated than both the simple and product group limits.  

In our construction, the lightest neutral $T$-odd particle
is expected to be the $\phi^0$ scalar.   (A neutral heavy gauge boson
might be the lightest neutral mode in extreme regions of parameter
space.)  However, it could be 
challenging to measure its spin especially if it is stable.  (See,
however, Ref.~\cite{Athanasiou:2006hv}.)  The mass of $\phi^0$ and its
couplings to Higgs are essentially free parameters in this theory as
discussed in Appendix C.  
As a result, it could have interesting consequences for Higgs
physics. For example, the Higgs could have a large invisible branching
ratio to such a scalar via a $h^{\dagger}h \phi^{\dagger} \phi$ coupling.

Next we consider the collider phenomenology away from the $T$-parity limit.
Notice that $T$-parity violation is probably only constrained by
precision electroweak measurements and flavor physics, and is
therefore not nearly as dangerous as lepton or baryon number violating
$R$-parity breaking in SUSY.  
Without $T$-parity, the new particles do not always need to be
pair-produced and there is no new stable neutral particle to give the
missing energy signature. 
They can be
searched for by looking for 
peaks in invariant mass distributions. For new gauge bosons, $Z_R$
and $W'_L$ in general have unsuppressed couplings to standard model
fermions and can be produced easily, dominated by the Drell-Yan
process. 
The couplings of the $X$ gauge
bosons to standard model fermions are suppressed by $v/f$, but LHC can
still have a significant reach ($\sim 2$ TeV) for
them~\cite{:1999fr}. On the other hand, discovering $W_R^\pm$ through
direct production in this model will be challenging as their couplings
to standard model fermions are suppressed by $v^2/f^2$.  It may be
more promising if they appear as decay products of other new
particles. Quark partners will have large QCD production cross
sections if they are not too heavy. In general they can decay to a
standard model quark and a new heavy gauge boson, which then
subsequently decay to standard model particles. If the $T$-parity
violations are small, the ratio of single and pair productions may
provide a measurement of the size of such violations. 
The decays of the approximate $T$-odd particles will mostly follow the decay chains of
the $T$-symmetric model until the last step, where the lightest
approximate $T$-odd particle decay into standard model particles. 

With the accumulation of higher luminosity, one can ask more detailed
questions. For example, what is the underlying global symmetry
structure that protects the Higgs mass? The moose presented here is
the minimal one that preserves custodial $SU(2)$ and in this 
sense is a good starting point for reconstructing the composite
symmetry structure. On the 
other hand, if we get more detailed information about the exotics, we
should be able to determine the global symmetry structure
precisely. For example, if there are fewer exotic gauge bosons, such
as no 
$W_R^{\pm}$, we might want to consider the global symmetry structure
considered in Section \ref{sec:toy}.  Alternatively, to account for
more scalar states, we could increase the size of the symmetry
breaking $G/H$ coset space.   

For the same reason, we will need to adjust the fermion structure
based on what we observe, in particular after we have some idea of
whether left- and/or right-handed partners of the standard model
fermions have been produced.  In the $T$-symmetric limit, it will be
important to study whether or not one can actually tell the difference
between left- and right-handed partner production at the LHC, and just
like in trying to distinguish between left- or right-handed squarks,
lepton production from cascade decays may be crucial in determining
whether or not the new partners have $SU(2)_L$
charges~\cite{Arkani-Hamed:2005px}.  If 
$T$-even partner fermions are accessible at the LHC, it may be easier
to determine their charges, spins, and couplings as their decays do
not necessarily yield large missing energy. Similar to the second KK
resonances in UED, detection of even
fermionic states, as resonances, could be also used as a clue which
distinguishes this scenario from $R$-parity conserving low energy
supersymmetry. 

Another interesting feature to pay attention to is the existence of
one or more gluon partners.  While composite models do not require any
heavy color octets, they are generically present in models with
extra-dimensional or holographic interpretations.  Though the
production cross section for gluon partners can be large, it may be
difficult to distinguish from partner fermion production without some
information about jet charges or lepton charge (a)symmetries.  This is
particularly true for the $T$-parity conserving case. 
If we observe gluon partners, it would then be necessary to add
additional structure to the moose, though if we ignore anomaly
cancellation, a KK-gluon can be accommodated in little M-theory by
introducing separate $SU(3)_C$ groups on each site with additional
link fields to break $SU(3)^3$ down to the diagonal. 

Notice that in order to go to the limits described in Section
\ref{sec:limits}, we need to take certain gauge couplings to be
strong. This means that in the parameter space of the little-M theory,  
there are regions where certain gauge modes are quite strongly coupled
and yet not completely decoupled. Such regions could be challenging to
simulate accurately with tree-level Monte Carlo tools, and one should
be careful making statements about the discovery reach for these
strongly coupled modes.  Note that we do not expect these strongly
coupled gauge bosons to form bound states with fermions, since the
mass of the gauge bosons also scale up with the coupling.   

Finally, away from the $T$-symmetric limit, little M-theory will mimic
a lot of the phenomenology of Higgsless theories as well.
Generically, one expects to see heavy gauge bosons at the LHC before
one sees the Higgs, and there may initially be some confusion about
whether the new spin-1 modes have the right couplings to unitarize
$W$-$W$ scattering.  If there is evidence for both new scalar and
vector particles with $SU(2)_L$ couplings, then we would have to
figure out a way to distinguish between a composite Higgs model with
vector partners and Higgsless theory with extra technipions.

\section{Outlook}

In this paper, we have used the fact that different ultraviolet
theories can yield the same low energy physics to develop a general
framework for describing non-SUSY physics at LHC energy scales.  The
$Sp(4)/SO(4)$ little M-theory interpolates between simple group-like
and minimal moose-like composite Higgs models, allowing for rich
collider phenomenology.  While the $Sp(4)/SO(4)$ moose is by no means
the unique choice for describing non-SUSY physics, it is a
well-motivated model that has the minimal symmetry structure
compatible with custodial $SU(2)$. 

Of course, at higher energies, different fundamental theories can be
distinguished from their M-theory approximations.  If a tower of
KK modes is seen whose masses fall at the roots of Bessel functions,
then a warped extra dimension (or a strongly coupled CFT) would be the
most straightforward explanation and a moose description would be
needlessly cumbersome.   Similarly, if KK gravitons are seen at the
LHC, then a moose description would be inappropriate, though there has been progress in developing healthy lattice descriptions of gravitational warped
dimensions \cite{Gallicchio:2005mh,Randall:2005me}.  However, if we
take the little hierarchy problem seriously, then we do not expect to
see strong dynamics or a plethora of KK states at the TeV scale.
Rather, we expect to find weakly coupled new physics, and theory space
is an especially convenient framework for describing new weakly
coupled non-SUSY physics. 

Though we have focused on composite Higgs models in this paper, with suitable modifications it is also possible to construct M-theories that interpolate between composite Higgs and UED theories.  Ignoring the Higgs sector, a little M-theory with $T$-parity could describe the
phenomenology of lower lying KK-modes in UED with KK-parity.  There are two important differences one would have to address to make this possible.   First, in UED there are same-statistics 
partners for every standard model field whereas in composite Higgs
theories there need not be the analog of the KK gluon. Additional
ingredients need to be added to little M-theory in order to describe those
states. Second, the KK fermions are Dirac in UED, so additional sites and
fermions need to be added to incorporate this fact.   Although
equivalent in principle, either an extra-dimensional or a moose description could be more
useful depending on the spectrum discovered at the LHC. If evenly spaced resonances nearly
degenerate in mass are discovered, UED will be undoubtedly be a much
simpler framework to work with. Otherwise, little M-theory type moose models would
be more useful since they allow for more general mass relations. 

It is also interesting that the variety of moose
models include Higgsless models and the low-lying resonances of
technicolor as well~\cite{Foadi:2003xa,Hirn:2004ze,Casalbuoni:2004id,Chivukula:2004pk,Perelstein:2004sc,Georgi:2004iy,SekharChivukula:2004mu,Chivukula:2006cg}. In both composite Higgs and
Higgsless theories, the longitudinal modes of the $W$ and $Z$ bosons
can be thought of as living in the $A_5$ component of a bulk gauge 
field.  Similarly, in the vector limit \cite{Georgi:1989xy}, the
$\rho$ meson and other light resonances of scaled-up QCD can be
described by multi-site mooses.  However, there are differences in
detailed realizations between composite Higgs and Higgsless theories
that make such an interpolation less useful. For example, in Higgsless
theories $W$-$W$ scattering is unitarized by a tower of spin-1 modes
instead of a spin-0 physical Higgs mode, so a master M-theory would
require both sets of 
unitarizing fields.   In addition, there is no useful notion of
$T$-parity in Higgsless theories, because $W$-$W$ scattering has to be
unitarized by tree-level exchange of spin-1 modes, so a $T$-symmetric
Higgsless theory would have additional $T$-odd states without
improving constraints from precision electroweak measurements.
Therefore, it is probably natural to think of composite Higgs and
Higgsless theories as two classes of moose models, just as the MSSM
and the NMSSM are two classes of SUSY models with different approaches
to the Higgs sector. 

Besides offering an interesting model for collider studies, little
M-theory also suggests an interesting philosophy for physics in the
LHC era.  Because there is no simple ultraviolet completion of the
$Sp(4)/SO(4)$ moose, little M-theory is unlikely to satisfy top-down
physicists who pine for UV complete models.  Because the $Sp(4)/SO(4)$
moose contains fixed relationships among some of the parameters,
little M-theory is unlikely to satisfy bottom-up physicists who would
rather measure Lagrangian parameters with no theoretical biases.
However, if there is a natural solution to the hierarchy problem and a
compelling explanation for the little hierarchy, then it is likely
that both top-down and bottom-up approaches will be necessary to
decipher LHC physics.  This is especially true if there is $T$-parity
and much of the decay topology information is lost as missing energy
at the LHC.  As a theoretically consistent model with a low 10 TeV
cutoff, little M-theory suggests a compromise between the top-down and
bottom-up approaches particularly well-suited for the LHC. 

\acknowledgments{We thank Nima Arkani-Hamed for useful
discussions, and especially Martin Schmaltz for advertising this model
before publication.  H.-C. C is supported by the Outstanding Junior
Investigator Award of the Department  of Energy.  J.T. is supported by
a fellowship from the Miller Institute for Basic Research in
Science. L.-T.~W is supported by the DOE under contract
DE-FG02-91ER40654.}

\appendix

\section{$Sp(4)$ Representations}

The elements of $Sp(4)$ consist of unitary matrices $P$ that satisfy
\be
P A P^T = A,
\ee
where $A$ is an anti-symmetric matrix.  In terms of generators $T$, every element of $Sp(4)$ can be written as $P = e^{i T \phi}$, where
\be
T A + A T^T = 0, \qquad \tr \left( T^a T^b \right) = \frac{1}{2} \delta^{ab}.
\ee
For convenience, we work in a basis where
\be
A = \left(\begin{array}{cccc}0 & 1 & 0 & 0 \\-1 & 0 & 0 & 0 \\0 & 0 & 0 & -1 \\0 & 0 & 1 & 0\end{array}\right),
\ee
and the 10 $Sp(4)$ generators that satisfy these conditions are
\be
T_{Li}=\frac{1}{2}\left(\begin{array}{cc}\sigma^i & 0 \\0 &
  0\end{array}\right), \quad
  T_{Ri}=\frac{1}{2}\left(\begin{array}{cc}0 & 0 \\0 & \sigma^i
  \end{array}\right), \quad
  T_{X0}=\frac{1}{2\sqrt{2}}\left(\begin{array}{cc}0 & 1 \\1 &
    0\end{array}\right), \quad
    T_{Xi}=\frac{1}{2\sqrt{2}}\left(\begin{array}{cc}0 & i \sigma^i \\
      -i \sigma^i & 0\end{array}\right). 
\ee
$T_{L,R}$ generate the $SU(2)_{L,R}$ subgroups of $Sp(4)$. 

\section{Mass Spectrum of the $Sp(4)/SO(4)$ Little M-theory}
\label{app:spectrum}

In this appendix, we discuss the mass spectrum of the $Sp(4)/SO(4)$ little M-theory.
In the gauge sector, before electroweak symmetry breaking, gauge bosons can be
classified according to their transformation properties under $SU(2)_L$ and $U(1)_{T_{3R}}$.
and there are mixings within each class. Only the middle site has $W_R^\pm$ and 
off-diagonal $X$ gauge bosons and their masses are given by
\be
M_{W_{mR}^{\pm}}^2 = M_{X_m}^2 = g_m^2 (f_1^2 + f_2^2). 
\ee
$W_{mR}^3$ and $W_R^3$ mix through the $\Sigma$ vevs and the mass matrix is 
\be
M_{(W_{mR}^3, W_R^3)}^2 = \left( \begin{array}{cc}
g_m^2 (f_1^2 + f_2^2)& -g_m g_R (f_1^2 + f_2^2)\\ -g_m g_R (f_1^2 + f_2^2)& g_R^2(f_1^2 + f_2^2) .
\end{array} \right)
\ee
The massless combination
\be
B=\frac{1}{\sqrt{g_R^2+g_m^2}}\left(g_R W_{mR}^3 + g_m W_{R}^3\right)
\ee
is identified with the hypercharge gauge boson of the standard model.
The standard model hypercharge coupling $g'$ is given by
\be
\frac{1}{g'^2}= \frac{1}{g_R^2}+\frac{1}{g_m^2}.
\ee
The orthogonal combination $Z_R$ acquires a mass-squared of
\be
(g_m^2+g_R^2) (f_1^2+ f_2^2).
\ee
There is a set of $SU(2)_L$ gauge bosons on each site. Their mass-squareds form
a $3\times 3$ matrix,
\be
\label{eq:wmassmatrix}
M_{(W_{1L}, W_{mL}, W_{2L})}^2 = \left( \begin{array}{ccc}
g_1^2 f_1^2 & - g_1 g_m f_1^2 & 0 \\ 
-g_1 g_m f_1^2 & g_m^2(f_1^2+f_2^2) & -g_2 g_m f_2^2 \\
0 & -g_2 g_m f_2^2  & g_2 f_2^2 
\end{array} \right).
\ee
After diagonalizing the matrices, there is one massless combination,
\be
W^{\pm,3}= \frac{1}{\sqrt{g_1^{-2}+g_m^{-2}+g_2^{-2}}}\left(\frac{W_{1L}^{\pm,3}}{g_1}+\frac{W_{mL}^{\pm,3}}{g_m}+\frac{W_{2L}^{\pm,3}}{g_2}\right)
\ee
which is identified with the standard model $W$ gauge bosons. 
The standard model $SU(2)_L$ gauge coupling $g$ is given by
\be
\frac{1}{g^2}= \frac{1}{g_1^2} +\frac{1}{g_m^2} +\frac{1}{g_2^2}.
\ee
There are two massive modes with mass-squareds of
\be
\nonumber
\frac{1}{2}\left( g_1^2 f_1^2 + g_2^2 f_2^2 + g_m^2 (f_1^2+f_2^2) \qquad \qquad \qquad \qquad \qquad \qquad \qquad \qquad \qquad \qquad \vphantom{\sqrt{(g_1^2)}}\right.
\ee
\be
\left. \qquad \qquad ~ \pm \sqrt{(g_1^2 f_1^2 + g_2^2 f_2^2 + g_m^2 (f_1^2+f_2^2))^2-4(g_1^2 g_m^2+g_2^2 g_m^2 +g_1^2 g_2^2) f_1^2 f_2^2}\right).
\ee
In the $T$-symmetric limit, $g_1=g_2=\bar{g},\, f_1=f_2=f= \sqrt{2} f_{\rm eff}$, they reduce to
\be
M^2_{W_{L,{\rm odd}}}=2\bar{g}^2 f_{\rm eff}^2, \quad M^2_{W_{L,{\rm even}}}=2(\bar{g}^2+2 g_m^2) f_{\rm eff}^2.
\ee

After electroweak symmetry breaking, $W^\pm$ and $Z=\cos\theta_W W^3 -\sin\theta_W B$ acquire masses and only
the photon $A= \sin\theta_W W^3 +\cos\theta_W B$ is left massless. The heavy gauge bosons (with mass $\sim gf$) also receive corrections from the electroweak symmetry breaking and there are further mixings among states with the same electric charge (and $T$-parity if it is a good symmetry). The corrections are small though (${\cal O}(v^2/f^2))$.

We now discuss the fermion mass spectrum and use the third generation quarks as an example. 
Before electroweak symmetry breaking, there are 3 $q$'s and 2 $q^c$'s for the doublets, so
one combination of $q_1,\,q_2$ and $q_m$ remains massless. Similarly,
there are 2 $t$'s and 3 $t^c$'s for the singlets so one combinations of
$t_1^c,\,t_2^c$ and $t'^c$ remains massless. They can be identified as the 
standard model top-bottom quark doublet and top quark singlet respectively.
They acquire a mass only after the electroweak symmetry breaking. The eigenstates
and eigenvalues are obtained by diagonalizing a $5\times 5$ mass matrix (ignoring
inter-generation mixings). The $5\times 5$ mass matrix to order ${\cal O}(v)$ is
given by
\bea
&  \hspace{20pt} q_1^c \hspace{45pt}  q_2^c \hspace{40pt}  t_1^c (b_1^c) \hspace{27pt}  t_2^c(b_2^c) \hspace{20pt}  t'^c (b'^c) \hspace{-16pt}  & \nonumber\\
&& \nonumber \\
\begin{matrix} q_1 \\ q_2 \\ q_m \\ t_m (b_m) \\ t'(b') \end{matrix} &\quad
\begin{pmatrix} m_1 & \alpha_1 f_{\rm eff} & 0 & \frac{i}{2\sqrt{2}}\alpha_1 v & 0 \\
\alpha_2 f_{\rm eff} & m_2 &  -\frac{i}{2\sqrt{2}}\alpha_2 v & 0 & 0\\
\beta_{q1} f_1 & \beta_{q2} f_2 & -\frac{i}{2\sqrt{2}}s\beta_{t(b)1}  v & \frac{i}{2\sqrt{2}}c\beta_{t(b)2}  v & 0\\
-\frac{i}{2\sqrt{2}}s\beta_{q1}  v & \frac{i}{2\sqrt{2}}c\beta_{q2}  v & \beta_{t(b)1} f_1 & \beta_{t(b)2} f_2 & 0\\
0 & 0 & F_{T(B)1} & -F_{T(B)2} & K_{T(B)}
\end{pmatrix} &,
\eea
where
\be
c= \frac{f_1}{\sqrt{f_1^2+f_2^2}}, \quad s= \frac{f_2}{\sqrt{f_1^2+f_2^2}},\quad f_{\rm eff} =s f_1= c f_2.
\ee

In the $T$-symmetric limit, $m_1=m_2=m,\, F_1=F_2=F,\, \beta_1=\beta_2=\beta,\,\alpha_1=\alpha_2=\alpha,\,s=c=1/\sqrt{2},\,f_1=f_2=f=\sqrt{2}f_{\rm eff}$, the even and odd states decouple. The mass matrix for the even states is
\bea
& \hspace{9pt} q_+^c \hspace{37pt} t_+^c(b_+^c) \hspace{17pt} t'^c(b'^c) \hspace{-10pt} & \nonumber \\
&& \nonumber \\
\begin{matrix} q_+ \\ q_m \\ t' \end{matrix} &
\begin{pmatrix} m + \alpha f_{\rm eff} & -\frac{i}{2\sqrt{2}}\alpha v & 0 \\
2\beta_q f_{\rm eff} & -\frac{i}{2\sqrt{2}}\beta_{t(b)} v & 0 \\
0 & \sqrt{2}F_{T(B)} & K_{T(B)} \end{pmatrix} & ,
\eea
and before electroweak symmetry breaking, the heavy masses are
\be
\sqrt{2F_{T(B)}^2 + K_{T(B)}^2}, \qquad \sqrt{(m + \alpha f_{\rm eff})^2 + 4 f_{\rm eff}^2 \beta_q^2}.
\ee
Apart from $\mathcal{O}(v^2/f^2)$ corrections, the standard model zero modes are
\be
\nonumber
q_{\rm SM} = \frac{1}{\sqrt{(m + \alpha f_{\rm eff})^2 + 4 f_{\rm eff}^2 \beta_q^2}}\left(  2 f_{\rm eff} \beta_q q_+ - (m + \alpha f_{\rm eff}) q_m \right),
\ee
\be
\qquad t^c_{\rm SM} = \frac{1}{\sqrt{K_T^2 + 2 F_T^2}} \left( K_T t^c_+ - \sqrt{2}F_T t'^c \right),
\ee
and the Yukawa coupling is
\be
\lambda_{\rm SM} = \frac{i K_{T(B)} (\beta_{t(b)}(m+\alpha f_{\rm eff}) - 2\beta_q \alpha f_{\rm eff})}{\sqrt{4(K_{T(B)}^2+2F_{T(B)}^2)(4\beta_q^2 f_{\rm eff}^2 + (m+\alpha f_{\rm eff})^2)}}.
\ee
In the $T$-odd sector, the fermion mass matrix is
\bea
& \hspace{6pt} q_-^c \hspace{33pt} t_-^c(b_-^c) \hspace{-10pt} & \nonumber \\
&& \nonumber \\
\begin{matrix} q_- \\ t_m(b_m) \end{matrix} &
\begin{pmatrix} m-\alpha f_{\rm eff} & \frac{i}{2\sqrt{2}}\alpha v \\
-\frac{i}{2\sqrt{2}}\beta_{q} v & 2\beta_{t(b)} f_{\rm eff} \end{pmatrix} & .
\eea

\section{Adjusting Goldstone Masses and Interactions}
\label{sec:goldstonemasses}

In this appendix, we discuss the potential and mass spectrum of the pseudo-Goldstone bosons, including the Higgs field and the extra scalars $\phi^{\pm,0}$.
As we mentioned above, how the Higgs potential arises in various little Higgs theories is very model-dependent. So, instead of specifying some particular mechanism to generate the Higgs potential, we will simply parametrize the scalar potential through appropriate operators made from the $\Sigma$ field. Non-trivial potentials for the Goldstones can be written down with insertions of symmetry-breaking spurions. In the custodial $SU(2)$ limit, the potential for $\Sigma$ can be written as
\bea
V(\Sigma) &=& f^4\bigg[ \kappa_1 \tr\Sigma\Theta\Sigma^\dagger\Theta +\kappa_2 \tr\Sigma\Theta\Sigma\Theta + \kappa_3 \tr\Sigma\Theta\Sigma\Theta\Sigma\Theta +\kappa_4 \tr\Sigma\Theta\Sigma\Theta\Sigma^\dagger\Theta \\ &&
+ \kappa_5 (\tr\Sigma\Theta\Sigma^\dagger\Theta)^2 + \kappa_6 \tr(\Sigma\Theta\Sigma^\dagger\Theta)^2 + \cdots + {\rm h.c.}
\bigg]
\eea
where $\Theta= (I-\Omega)/2 = {\rm diag}(0,0,1,1)$ is the spurion matrix for the breaking of $Sp(4)$. $\Theta$ can be inserted between two $\Sigma$'s because there is only one common $U(1)_R$ gauged in the bottom two rows and columns. Each of these terms contains various combinations of masses and interactions of the Goldstones, for instance,
\bea
f^4\, {\rm tr}\,\Sigma\, \Theta\, \Sigma^\dagger\, \Theta &= &   f^4 \left( 2 - \frac{ h^\dagger h }{2f^2}+\cdots \right), \nonumber \\
f^4\, {\rm tr}\, \Sigma\, \Theta\, \Sigma\, \Theta &= &  f^4 \left(2 - \frac{ h^\dagger h }{2f^2} - 2\frac{\phi^+ \phi^-}{f^2} - \frac{(\phi^0)^2}{f^2} +\cdots \right).
\eea 
Therefore, by choosing the coefficients $\kappa_1,\, \kappa_2, \cdots$, one can produce any scalar potential consistent with the symmetries. Under the $SU(2)_L$ (gauged) and $SU(2)_R$ (custodial, with the $U(1)_R$ subgroup gauged), the Higgs multiplet
\be
H = \frac{1}{\sqrt{2}}\left( \begin{array}{cc} h^0 & h^- \\ -h^+ & h^{0*}\end{array}\right)
\ee
transforms as ({\bf 2, 2}) and the remaining Goldstones
\be
\Phi=\left( \begin{array}{cc} \phi^0/2  & \phi^-/\sqrt{2} \\ \phi^+/\sqrt{2} & -\phi^{0}/2 \end{array}\right)
\ee
transforms as ({\bf 1, 3}).
The leading scalar potential can then be written as
\be
\label{eq:scalarpotential}
V(H,\Phi) =  m_H^2 \tr H^\dagger H + m_\Phi^2 \tr \Phi^2 + \frac{\lambda_H}{2} (\tr H^\dagger H)^2 + \frac{\lambda_\Phi}{2} (\tr \Phi^2)^2 + \lambda_{H\Phi} \tr\Phi^2 \tr H^\dagger H + \cdots.
\ee
Note that there is no trilinear term (in the custodial $SU(2)$ limit) because $H$ always appears in the singlet combination $\tr H^\dagger H$ due to the antisymmetric property of the $SU(2)$ $\epsilon$ tensor. We take $m_H^2<0$ so that electroweak symmetry is broken correctly. Some tuning on the parameters is required to get $v_{\rm EW} \ll f$ as we do not specify the origin of these parameters.

The degeneracy between $\phi^{\pm}$ and $\phi^0$ will be lifted by the custodial $SU(2)$ violating effects, including the radiative corrections coming from the $U(1)_R$ gauge field. The custodial $SU(2)$ breaking can be parametrized by the
spurion matrix $\Xi= {\rm diag}(0,0,1,-1)$, for instance,
\bea
f^4\, {\rm tr}\, \Sigma\, \Xi\, \Sigma^\dagger\, \Xi\, &= & f^4 \left(2- 2\frac{\phi^+ \phi^-}{f^2} -\frac{ h^\dagger h}{2f^2} +\cdots\right), \nonumber \\
\label{eq:plaquettes}
f^4\, {\rm tr}\, \Sigma\, \Xi\, \Sigma\, \Xi\, &= & f^4 \left(2- \frac{(\phi^0)^2}{f^2}-\frac{ h^\dagger h}{2f^2} + \cdots \right).
\eea
These effects can be thought of as $\phi^0$ having a nonzero vacuum expectation value. (Indeed $\phi^0$ can have a tadpole term in the presence of custodial $SU(2)$ breaking.) The possible scalar potential can be obtained by shifting $\phi^0$ in \eq{eq:scalarpotential} by a constant. A trilinear term
\be
\phi^0 \tr H^\dagger H
\ee 
 and other terms linear in $\phi^0$ are now possible, but will be suppressed if the custodial $SU(2)$ breaking effects are small.


\begin{thebibliography}{99}

\bibitem{Weinberg:1975gm}
S.~Weinberg,
\emph{Implications Of Dynamical Symmetry Breaking,}
Phys.\ Rev.\ D {\bf 13}, 974 (1976).

\bibitem{Susskind:1978ms}
L.~Susskind,
\emph{Dynamics Of Spontaneous Symmetry Breaking In The Weinberg-Salam Theory,}
Phys.\ Rev.\ D {\bf 20}, 2619 (1979).


\bibitem{Hill:1991at}
  C.~T.~Hill,
  \emph{Topcolor: Top quark condensation in a gauge extension of the standard model,}
  Phys.\ Lett.\ B {\bf 266}, 419 (1991).

\bibitem{Csaki:2003dt}
  C.~Csaki, C.~Grojean, H.~Murayama, L.~Pilo and J.~Terning,
  \emph{Gauge theories on an interval: Unitarity without a Higgs,}
  Phys.\ Rev.\ D {\bf 69}, 055006 (2004)
  [arXiv:hep-ph/0305237].
  
  
\bibitem{Arkani-Hamed:2000hv}
  N.~Arkani-Hamed, H.~C.~Cheng, B.~A.~Dobrescu and L.~J.~Hall,
  \emph{Self-breaking of the standard model gauge symmetry,}
  Phys.\ Rev.\ D {\bf 62}, 096006 (2000)
  [arXiv:hep-ph/0006238].

\bibitem{Appelquist:2000nn}
  T.~Appelquist, H.~C.~Cheng and B.~A.~Dobrescu,
  \emph{Bounds on universal extra dimensions,}
  Phys.\ Rev.\ D {\bf 64}, 035002 (2001)
  [arXiv:hep-ph/0012100].

\bibitem{Arkani-Hamed:2001nc}
N.~Arkani-Hamed, A.~G.~Cohen and H.~Georgi,
\emph{Electroweak symmetry breaking from dimensional deconstruction,}
Phys.\ Lett.\ B {\bf 513}, 232 (2001)
[arXiv:hep-ph/0105239].

\bibitem{Arkani-Hamed:2002pa}
N.~Arkani-Hamed, A.~G.~Cohen, T.~Gregoire and J.~G.~Wacker,
\emph{Phenomenology of electroweak symmetry breaking from theory space,}
JHEP {\bf 0208}, 020 (2002)
[arXiv:hep-ph/0202089].

\bibitem{Contino:2003ve}
  R.~Contino, Y.~Nomura and A.~Pomarol,
  \emph{Higgs as a holographic pseudo-Goldstone boson,}
  Nucl.\ Phys.\ B {\bf 671}, 148 (2003)
  [arXiv:hep-ph/0306259].
 
  
\bibitem{Agashe:2004rs}
  K.~Agashe, R.~Contino and A.~Pomarol,
  \emph{The minimal composite Higgs model,}
  Nucl.\ Phys.\ B {\bf 719}, 165 (2005)
  [arXiv:hep-ph/0412089].

\bibitem{Chacko:2005pe}
  Z.~Chacko, H.~S.~Goh and R.~Harnik,
  \emph{The twin Higgs: Natural electroweak breaking from mirror symmetry,}
  arXiv:hep-ph/0506256.


\bibitem{Dimopoulos:1981zb}
  S.~Dimopoulos and H.~Georgi,
  \emph{Softly Broken Supersymmetry And SU(5),}
  Nucl.\ Phys.\ B {\bf 193}, 150 (1981).

\bibitem{Georgi:1985hf}
  H.~Georgi,
  \emph{A Tool Kit For Builders Of Composite Models,}
  Nucl.\ Phys.\ B {\bf 266}, 274 (1986).

\bibitem{ADD}
  N.~Arkani-Hamed, S.~Dimopoulos and G.~R.~Dvali,
  \emph{The hierarchy problem and new dimensions at a millimeter,}
  Phys.\ Lett.\ B {\bf 429}, 263 (1998)
  [arXiv:hep-ph/9803315].

\bibitem{RS1}
  L.~Randall and R.~Sundrum,
  \emph{A large mass hierarchy from a small extra dimension,}
  Phys.\ Rev.\ Lett.\  {\bf 83}, 3370 (1999)
  [arXiv:hep-ph/9905221].
  
\bibitem{RS2}
L.~Randall and R.~Sundrum,
  \emph{An alternative to compactification,}
  Phys.\ Rev.\ Lett.\  {\bf 83}, 4690 (1999)
  [arXiv:hep-th/9906064].

\bibitem{Kaplan:2003uc}
  D.~E.~Kaplan and M.~Schmaltz,
  \emph{The little Higgs from a simple group,}
  JHEP {\bf 0310}, 039 (2003)
  [arXiv:hep-ph/0302049].

\bibitem{Schmaltz:2004de}
  M.~Schmaltz,
  \emph{The simplest little Higgs,}
  JHEP {\bf 0408}, 056 (2004)
  [arXiv:hep-ph/0407143].

\bibitem{Arkani-Hamed:2002qx}
  N.~Arkani-Hamed, A.~G.~Cohen, E.~Katz, A.~E.~Nelson, T.~Gregoire and J.~G.~Wacker,
  \emph{The minimal moose for a little Higgs,}
  JHEP {\bf 0208}, 021 (2002)
  [arXiv:hep-ph/0206020].

\bibitem{Chang:2003un}
  S.~Chang and J.~G.~Wacker,
  \emph{Little Higgs and custodial SU(2),}
  Phys.\ Rev.\ D {\bf 69}, 035002 (2004)
  [arXiv:hep-ph/0303001].

\bibitem{Maldacena:1997re}
J.~M.~Maldacena,
\emph{The large N limit of superconformal field theories and supergravity,}
Adv.\ Theor.\ Math.\ Phys.\  {\bf 2}, 231 (1998)
[Int.\ J.\ Theor.\ Phys.\  {\bf 38}, 1113 (1999)]
[arXiv:hep-th/9711200].

\bibitem{Gubser:1998bc}
S.~S.~Gubser, I.~R.~Klebanov and A.~M.~Polyakov,
\emph{Gauge theory correlators from non-critical string theory,}
Phys.\ Lett.\ B {\bf 428}, 105 (1998)
[arXiv:hep-th/9802109].

\bibitem{Witten:1998qj}
E.~Witten,
\emph{Anti-de Sitter space and holography,}
Adv.\ Theor.\ Math.\ Phys.\  {\bf 2}, 253 (1998)
[arXiv:hep-th/9802150].

\bibitem{Arkani-Hamed:2000ds}
N.~Arkani-Hamed, M.~Porrati and L.~Randall,
\emph{Holography and phenomenology,}
JHEP {\bf 0108}, 017 (2001)
[arXiv:hep-th/0012148].

\bibitem{Rattazzi:2000hs}
R.~Rattazzi and A.~Zaffaroni,
\emph{Comments on the holographic picture of the Randall-Sundrum model,}
JHEP {\bf 0104}, 021 (2001)
[arXiv:hep-th/0012248].

\bibitem{Thaler:2005kr}
  J.~Thaler,
  \emph{Little technicolor,}
  JHEP {\bf 0507}, 024 (2005)
  [arXiv:hep-ph/0502175].

\bibitem{Arkani-Hamed:2001ca}
  N.~Arkani-Hamed, A.~G.~Cohen and H.~Georgi,
  \emph{(De)constructing dimensions,}
  Phys.\ Rev.\ Lett.\  {\bf 86}, 4757 (2001)
  [arXiv:hep-th/0104005].

\bibitem{Hill:2000mu}
  C.~T.~Hill, S.~Pokorski and J.~Wang,
  \emph{Gauge invariant effective Lagrangian for Kaluza-Klein modes,}
  Phys.\ Rev.\ D {\bf 64}, 105005 (2001)
  [arXiv:hep-th/0104035].
  
\bibitem{Barbieri:2000gf}
  R.~Barbieri and A.~Strumia,
  \emph{The 'LEP paradox',}
  arXiv:hep-ph/0007265.

\bibitem{Barbieri:2004qk}
  R.~Barbieri, A.~Pomarol, R.~Rattazzi and A.~Strumia,
  \emph{Electroweak symmetry breaking after LEP1 and LEP2,}
  Nucl.\ Phys.\ B {\bf 703}, 127 (2004)
  [arXiv:hep-ph/0405040].

\bibitem{Cheng:2003ju}
  H.~C.~Cheng and I.~Low,
  \emph{TeV symmetry and the little hierarchy problem,}
  JHEP {\bf 0309}, 051 (2003)
  [arXiv:hep-ph/0308199].

\bibitem{Cheng:2004yc}
  H.~C.~Cheng and I.~Low,
  \emph{Little hierarchy, little Higgses, and a little symmetry,}
  JHEP {\bf 0408}, 061 (2004)
  [arXiv:hep-ph/0405243].

\bibitem{Kaplan:1983fs}
D.~B.~Kaplan and H.~Georgi,
\emph{SU(2) X U(1) Breaking By Vacuum Misalignment,}
Phys.\ Lett.\ B {\bf 136}, 183 (1984).

\bibitem{Kaplan:1983sm}
D.~B.~Kaplan, H.~Georgi and S.~Dimopoulos,
\emph{Composite Higgs Scalars,}
Phys.\ Lett.\ B {\bf 136}, 187 (1984).

\bibitem{Arkani-Hamed:2002qy}
  N.~Arkani-Hamed, A.~G.~Cohen, E.~Katz and A.~E.~Nelson,
  \emph{The littlest Higgs,}
  JHEP {\bf 0207}, 034 (2002)
  [arXiv:hep-ph/0206021].

\bibitem{Schwarz:1995jq}
  J.~H.~Schwarz,
  \emph{The power of M theory,}
  Phys.\ Lett.\ B {\bf 367}, 97 (1996)
  [arXiv:hep-th/9510086].

\bibitem{Cheng:2001nh}
  H.~C.~Cheng, C.~T.~Hill and J.~Wang,
  \emph{Dynamical electroweak breaking and latticized extra dimensions,}
  Phys.\ Rev.\ D {\bf 64}, 095003 (2001)
  [arXiv:hep-ph/0105323].

\bibitem{Randall:2002qr}
  L.~Randall, Y.~Shadmi and N.~Weiner,
  \emph{Deconstruction and gauge theories in AdS(5),}
  JHEP {\bf 0301}, 055 (2003)
  [arXiv:hep-th/0208120].

\bibitem{Luty:2004ye}
  M.~A.~Luty and T.~Okui,
  \emph{Conformal technicolor,}
  arXiv:hep-ph/0409274.

\bibitem{Agashe:2005dk}
  K.~Agashe and R.~Contino,
  \emph{The minimal composite Higgs model and electroweak precision tests,}
  Nucl.\ Phys.\ B {\bf 742}, 59 (2006)
  [arXiv:hep-ph/0510164].

\bibitem{Agashe:2006at}
  K.~Agashe, R.~Contino, L.~Da Rold and A.~Pomarol,
  \emph{A custodial symmetry for Z b anti-b,}
  arXiv:hep-ph/0605341.


\bibitem{Coleman:1969sm}
S.~R.~Coleman, J.~Wess and B.~Zumino,
\emph{Structure Of Phenomenological lagrangians. 1,}
Phys.\ Rev.\  {\bf 177}, 2239 (1969).

\bibitem{Callan:1969sn}
C.~G.~Callan, S.~R.~Coleman, J.~Wess and B.~Zumino,
\emph{Structure Of Phenomenological lagrangians. 2,}
Phys.\ Rev.\  {\bf 177}, 2247 (1969).

\bibitem{Bando:1987br}
M.~Bando, T.~Kugo and K.~Yamawaki,
\emph{Nonlinear Realization And Hidden Local Symmetries,}
Phys.\ Rept.\  {\bf 164}, 217 (1988).

\bibitem{Abbott:1981re}
L.~F.~Abbott and E.~Farhi,
\emph{Are The Weak Interactions Strong?,}
Phys.\ Lett.\ B {\bf 101}, 69 (1981).

\bibitem{Abbott:1981yg}
L.~F.~Abbott and E.~Farhi,
\emph{A Confining Model Of The Weak Interactions,}
Nucl.\ Phys.\ B {\bf 189}, 547 (1981).

\bibitem{Cacciapaglia:2004rb}
  G.~Cacciapaglia, C.~Csaki, C.~Grojean and J.~Terning,
  \emph{Curing the ills of Higgsless models: The S parameter and unitarity,}
  Phys.\ Rev.\ D {\bf 71}, 035015 (2005)
  [arXiv:hep-ph/0409126].
  
\bibitem{Foadi:2004ps}
  R.~Foadi, S.~Gopalakrishna and C.~Schmidt,
  \emph{Effects of fermion localization in Higgsless theories and electroweak
  constraints,}
  Phys.\ Lett.\ B {\bf 606}, 157 (2005)
  [arXiv:hep-ph/0409266].

\bibitem{Arkani-Hamed:2002sp}
  N.~Arkani-Hamed, H.~Georgi and M.~D.~Schwartz,
  \emph{Effective field theory for massive gravitons and gravity in theory space,}
  Annals Phys.\  {\bf 305}, 96 (2003)
  [arXiv:hep-th/0210184].

\bibitem{Arkani-Hamed:2003vb}
  N.~Arkani-Hamed and M.~D.~Schwartz,
  \emph{Discrete gravitational dimensions,}
  Phys.\ Rev.\ D {\bf 69}, 104001 (2004)
  [arXiv:hep-th/0302110].

\bibitem{Thaler:2005en}
  J.~Thaler and I.~Yavin,
  \emph{The littlest Higgs in anti-de Sitter space,}
  JHEP {\bf 0508}, 022 (2005)
  [arXiv:hep-ph/0501036].

\bibitem{Han:2003wu}
  T.~Han, H.~E.~Logan, B.~McElrath and L.~T.~Wang,
  \emph{Phenomenology of the little Higgs model,}
  Phys.\ Rev.\ D {\bf 67}, 095004 (2003)
  [arXiv:hep-ph/0301040].

\bibitem{Csaki:2002qg}
  C.~Csaki, J.~Hubisz, G.~D.~Kribs, P.~Meade and J.~Terning,
  \emph{Big corrections from a little Higgs,}
  Phys.\ Rev.\ D {\bf 67}, 115002 (2003)
  [arXiv:hep-ph/0211124].

\bibitem{Hewett:2002px}
  J.~L.~Hewett, F.~J.~Petriello and T.~G.~Rizzo,
  \emph{Constraining the littlest Higgs,}
  JHEP {\bf 0310}, 062 (2003)
  [arXiv:hep-ph/0211218].


\bibitem{Marandella:2005wd}
  G.~Marandella, C.~Schappacher and A.~Strumia,
  \emph{Little-Higgs corrections to precision data after LEP2,}
  Phys.\ Rev.\ D {\bf 72}, 035014 (2005)
  [arXiv:hep-ph/0502096].

\bibitem{Meade:2006dw}
  P.~Meade and M.~Reece,
  \emph{Top partners at the LHC: Spin and mass measurement,}
  arXiv:hep-ph/0601124.

\bibitem{isospinviolation}
 The LEP Collaborations ALEPH, DELPHI, L3, OPAL and the LEP
Electroweak Working Group and the SLD Heavy Flavor Group, \emph{A
combination of preliminary electroweak measurements and constraints
on the standard model,} arXiv:hep-ex/0212036.

\bibitem{Low:2004xc}
  I.~Low,
  \emph{T parity and the littlest Higgs,}
  JHEP {\bf 0410}, 067 (2004)
  [arXiv:hep-ph/0409025].

\bibitem{Arkani-Hamed:2005px}
  N.~Arkani-Hamed, G.~L.~Kane, J.~Thaler and L.~T.~Wang,
  \emph{Supersymmetry and the LHC inverse problem,}
  arXiv:hep-ph/0512190.

\bibitem{Athanasiou:2006hv}
  C.~Athanasiou, C.~G.~Lester, J.~M.~Smillie and B.~R.~Webber,
  \emph{Addendum to 'Distinguishing spins in decay chains at the Large Hadron
  Collider','}
  arXiv:hep-ph/0606212.
 
\bibitem{:1999fr}
 \emph{ATLAS detector and physics performance. Technical design report.  Vol. 2,}
CERN-LHCC-99-15


\bibitem{Gallicchio:2005mh}
  J.~Gallicchio and I.~Yavin,
  \emph{Curvature as a remedy or discretizing gravity in warped dimensions,}
  arXiv:hep-th/0507105.

\bibitem{Randall:2005me}
  L.~Randall, M.~D.~Schwartz and S.~Thambyapillai,
  \emph{Discretizing gravity in warped spacetime,}
  JHEP {\bf 0510}, 110 (2005)
  [arXiv:hep-th/0507102].

\bibitem{Foadi:2003xa}
  R.~Foadi, S.~Gopalakrishna and C.~Schmidt,
  \emph{Higgsless electroweak symmetry breaking from theory space,}
  JHEP {\bf 0403}, 042 (2004)
  [arXiv:hep-ph/0312324].

\bibitem{Hirn:2004ze}
  J.~Hirn and J.~Stern,
  \emph{The role of spurions in Higgs-less electroweak effective theories,}
  Eur.\ Phys.\ J.\ C {\bf 34}, 447 (2004)
  [arXiv:hep-ph/0401032].
  
\bibitem{Casalbuoni:2004id}
  R.~Casalbuoni, S.~De Curtis and D.~Dominici,
  \emph{Moose Models With Vanishing S Parameter,}
  Phys.\ Rev.\ D {\bf 70}, 055010 (2004)
  [arXiv:hep-ph/0405188].

\bibitem{Chivukula:2004pk}
  R.~S.~Chivukula, E.~H.~Simmons, H.~J.~He, M.~Kurachi and M.~Tanabashi,
   \emph{The structure of corrections to electroweak interactions in Higgsless
  models,}
  Phys.\ Rev.\ D {\bf 70}, 075008 (2004)
  [arXiv:hep-ph/0406077].

\bibitem{Perelstein:2004sc}
  M.~Perelstein,
  \emph{Gauge-assisted technicolor?,}
  JHEP {\bf 0410}, 010 (2004)
  [arXiv:hep-ph/0408072].

\bibitem{Georgi:2004iy}
  H.~Georgi,
  \emph{Fun with Higgsless theories,}
  Phys.\ Rev.\ D {\bf 71}, 015016 (2005)
  [arXiv:hep-ph/0408067].
  
\bibitem{SekharChivukula:2004mu}
  R.~Sekhar Chivukula, E.~H.~Simmons, H.~J.~He, M.~Kurachi and M.~Tanabashi,
  \emph{Electroweak corrections and unitarity in linear moose models,}
  Phys.\ Rev.\ D {\bf 71}, 035007 (2005)
  [arXiv:hep-ph/0410154].

\bibitem{Chivukula:2006cg}
  R.~S.~Chivukula, B.~Coleppa, S.~Di Chiara, H.~J.~He, M.~Kurachi, E.~H.~Simmons and M.~Tanabashi,
  \emph{A Three Site Higgsless Model,'}
  arXiv:hep-ph/0607124.
  
\bibitem{Georgi:1989xy}
H.~Georgi,
\emph{Vector Realization Of Chiral Symmetry,}
Nucl.\ Phys.\ B {\bf 331}, 311 (1990).






\end{thebibliography}
\end{document}